\documentclass[review]{elsarticle}
\pdfoutput=1

\usepackage{hyperref}

\usepackage{amsmath}
\usepackage{siunitx}
\usepackage{booktabs}

\usepackage{import}
\usepackage{xcolor}

\journal{Journal of Computational Physics}

\begin{document}

\begin{frontmatter}

\title{Multi-Objective CFD-Driven Development of Coupled Turbulence Closure Models}

\author[melbourne]{Fabian Waschkowski\corref{correspondingauthor}}
\cortext[correspondingauthor]{Corresponding author}
\ead{fwaschkowski@student.unimelb.edu.au}

\author[beijing]{Yaomin Zhao}
\author[melbourne]{Richard Sandberg}
\author[melbourne]{Joseph Klewicki}

\address[melbourne]{Department of Mechanical Engineering, University of Melbourne, VIC 3010, Australia}
\address[beijing]{Center for Applied Physics and Technology, HEDPS, College of Engineering, Peking University, Beijing 100871, China}

\begin{abstract}
This paper introduces two novel concepts in data-driven turbulence modeling that enable the simultaneous development of multiple closure models and the training towards multiple objectives. The concepts extend the evolutionary framework by Weatheritt and Sandberg (2016) \cite{weatheritt2016}, which derives interpretable and implementation-ready expressions from high-fidelity simulation data. By assigning a shared fitness value to the evolved closure models and utilizing the CFD-driven training approach by Zhao et al. (2020) \cite{zhao2020}, the multi-expression training concept introduced here is able to account for the coupling between the trained models, i.e. Reynolds stress anisotropy, turbulent heat flux and turbulence production correction models. As a second concept, a multi-objective optimization algorithm is applied to the framework. The extension yields a diverse set of candidate models and allows a trade-off between the training objectives after analyzing the training results. In this study, the novel concepts are applied to a benchmark periodic hills case and a vertical natural convection flow. The predictions of mean flow quantities are improved compared to decoupled training strategies with distinct and robust improvements for strongly coupled momentum and thermal fields. The coupled training of closure models and the balancing of multiple training objectives are considered important capabilities on the path towards generalized data-driven turbulence models.

\end{abstract}


\end{frontmatter}

\section{Introduction}
\label{sec:introduction}

Reynolds-Averaged Navier-Stokes (RANS) calculations are the primary tool to perform Computational Fluid Dynamics (CFD) simulations in industrial applications. Despite a significant increase in computational resources over the past decades, high-fidelity simulation techniques, such as Direct Numerical Simulations (DNS) or Large Eddy Simulations (LES), remain computationally impractical for typical iterative design processes in industry. While RANS calculations are more affordable, the Reynolds averaging procedure introduces additional terms to the governing transport equations, namely the Reynolds stress tensor and the turbulent scalar flux vector. The two terms describe the impact of the turbulent fluctuations on the mean flow quantities and require closure models. The accuracy of RANS predictions depends on the choice of these closures. 

The most frequently applied turbulence models rely on the linear eddy-viscosity concept and a gradient diffusion hypothesis to model the Reynolds stresses and the turbulent scalar fluxes, respectively. While the simplicity and numerical robustness of these two approaches are reasons for their popularity, their limitations to accurately predict flows that include separation, adverse pressure gradients, curvature or buoyancy effects are widely known \cite{leschziner2015}.

The success of machine learning (ML) techniques in areas such as computer vision \cite{krizhevsky2012} and natural language processing \cite{vaswani2017} have inspired the turbulence modeling community to develop advanced closure models by utilizing data-driven algorithms and high-fidelity simulation data \cite{duraisamy2019}. \citet{cheung2011} and \citet{edeling2014a} applied Bayesian statistics to calibrate the model coefficients of different linear eddy-viscosity models (LEVM) for turbulent boundary layer flows. \citet{tracey2015} trained a shallow artificial neural network (ANN) to rediscover the source terms of the Spalart-Allmaras model. Several authors, such as \citet{duraisamy2015, parish2016} and \citet{holland2019}, used Bayesian inversion to derive scalar correction fields for different terms of the turbulence transport equations. In a second step, the spatially-dependent correction fields were then reconstructed as functions of mean flow quantities using ANNs or Gaussian processes. To overcome the limitations of the linear eddy-viscosity concept, i.e. the linear dependency of the Reynolds stress anisotropy on the mean strain rate, random forest models were trained to directly predict the anisotropy tensor \cite{ling2016,wu2018,kaandorp2020}. \citet{ling2016a} introduced a deep ANN architecture that incorporates an integrity basis of the anisotropy tensor

\begin{equation}
a_{ij} = \sum\limits_{k=1}^{10} g^k\left(I^1, I^2, ..., I^5\right)V_{ij}^k, 
\label{eq:integrity_basis}
\end{equation}

\noindent in which $a_{ij}$ is a linear combination of ten basis tensors $V_{ij}^k$ and depends on five scalar invariants $I^\lambda$. Both $V_{ij}^k$ and $I^\lambda$ are calculated from the non-dimensionalized mean strain and rotation rate tensors \cite{pope1975}. \citet{milani2021} extended this framework to model turbulent scalar fluxes using a vector basis proposed by \citet{zheng1994}.

While the above reviewed publications demonstrate promising reductions of the values of their respective training objectives, the applied algorithms have in common that their resulting models are not interpretable and a direct integration into a CFD solver is numerically challenging, which limits the utility for industrial applications. In order to produce implementable closure models, \citet{weatheritt2016} developed a framework based on Gene Expression Programming (GEP), which is a specific type of evolutionary algorithm \cite{ferreira2001}. The framework is termed EVE (EVolutionary algorithm for the development of Expressions) and symbolically regresses tangible algebraic equations, referred to as expressions, from high-fidelity data. The EVE framework was successfully applied to minimize errors in the prediction of the anisotropy tensor and the turbulent heat flux vector for various flow problems, such as separated flows, wake mixing and wall jets \cite{weatheritt2016,sandberg2018,akolekar2019}. A different framework to derive implementation-ready models was developed by \citet{schmelzer2019}. This method performs symbolic regression on a library of candidate equation snippets using a sparsity-enforcing regression technique. The authors applied the framework to correct the anisotropy prediction and the production term in the transport equations of the $k$-$\omega$ SST model.

An issue that arises in the practical application of machine-learned Reynolds stress anisotropy models is that even very small errors in the anisotropy prediction can yield unsatisfactory predictions of mean flow quantities, such as velocity, temperature or pressure, when solving the RANS transport equations \cite{thompson2016}. \citet{schmelzer2019} approached this problem by not deciding on one specific model during the regression step, but outputting a pool of candidate models that are ultimately evaluated in a RANS calculation. In order to address this problem using the EVE framework, \citet{zhao2020} introduced the concept of CFD-driven training, where a RANS calculation is performed for each candidate model throughout the training process. While this training approach is computationally more expensive, the impact of model prediction errors on the mean flow quantities is taken into account in the training process. An additional benefit is that any quantity available in the RANS results and the high-fidelity training data can be selected as a training target. This enables a variety of new options in the formulation of training objectives.

A second issue that is present in all current frameworks for the data-driven development of closure models is that only a single model can be derived per training run. When multiple models are developed, separate training runs are performed. Thus, the coupling of the models and their coupled effects on the RANS predictions are not considered. For example, when two models for the anisotropy and the turbulent heat flux are derived, the resulting RANS prediction of the mean temperature depends on the turbulence heat flux model and the mean velocity field. On the other hand, the mean velocity depends on the anisotropy model and, depending on the underlying assumptions, on the mean temperature. Not considering these coupling effects in the training process can lead to significant modeling errors for strongly coupled models and flows.

Lastly, there is no published framework for turbulence modeling that allows a training towards multiple objectives. To the authors' knowledge, all current frameworks implement a single training objective. When multiple objectives are of interest, the different objective functions are combined into a single composite function, which limits the exploration of the objective function space significantly \cite{konak2006}. The training of models for different flow problems (e.g. varied domains with similar underlying flow physics), for different characteristic parameters (e.g. a problem-specific range of Reynolds numbers), or to predict different mean flow quantities could benefit from distinct training objectives.

In this paper, we approach the described issues by introducing two concepts novel to data-driven turbulence modeling, namely multi-expression training and multi-objective optimization. The first concept develops expressions, i.e. tangible algebraic equations, for multiple closure models simultaneously based on the EVE framework \cite{weatheritt2016} and the CFD-driven training approach \cite{zhao2020}. Importantly, since the candidate models are evaluated using RANS calculations during the training process, the coupled effect of the models is considered in the training objective. The details of multi-expression training are discussed in Section \ref{sec:multi-expression}. Furthermore, the EVE framework is extended to perform multi-objective optimization instead of the existing training towards a single objective. This second concept is introduced in Section \ref{sec:multi-objective}.

The novel capabilities of the EVE framework are applied to a frequently used periodic hills benchmark case and a vertical natural convection flow. Multiple closure models are developed for each of the two canonical flow cases. The details of the flows, the closure models and the respective modeling strategies are presented in Section \ref{sec:strategy}. The results of training and testing the models are discussed in Section \ref{sec:results} and conclusions are drawn in Section \ref{sec:conclusion}.

\section{Methodology}
\label{sec:methodology}

In this section, we provide first a brief overview of the EVE framework as developed by \citet{weatheritt2016} and \citet{zhao2020}. Afterwards, we introduce the applied modifications that allow the framework to perform multi-expression training and multi-objective optimization in Sections \ref{sec:multi-expression} and \ref{sec:multi-objective}, respectively.

\subsection{Standard EVE Framework}
\label{sec:standard}

\begin{figure}[t]
    \centering
    \includegraphics{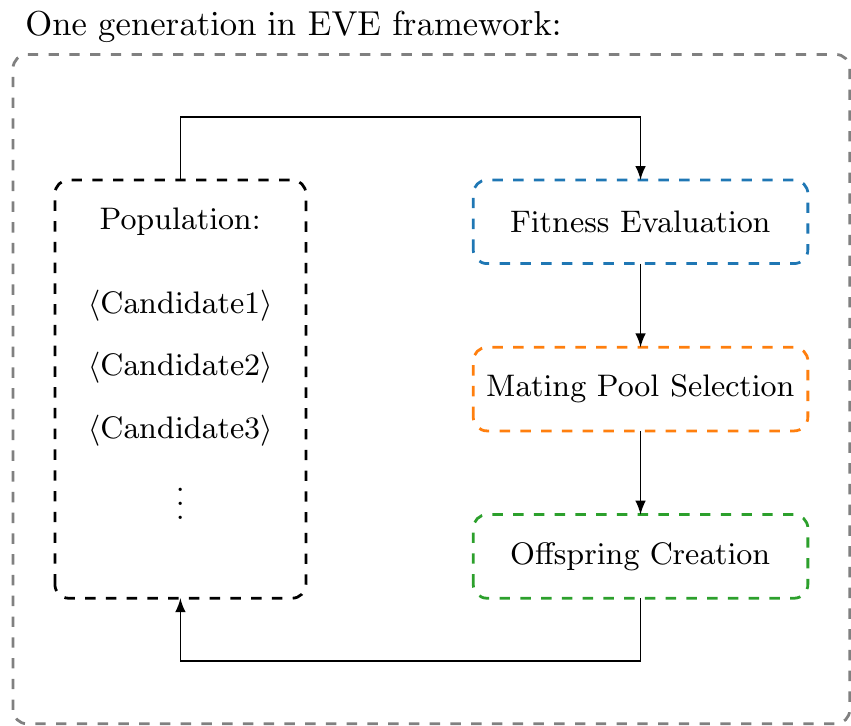}
    \caption{Flow chart of EVE framework \citep{weatheritt2016}.}
	\label{fig:generation}
\end{figure}

The EVE framework implements the GEP algorithm by \citet{ferreira2001}, which evolves a population of candidate solutions over various generations towards a specified training objective. For example, in the case of training a Reynolds stress anisotropy model, each candidate solution is one algebraic equation for the anisotropy tensor and the training objective could be to minimize the mean squared prediction error to the anisotropy data from a high-fidelity simulation. As such, an initial population is created randomly and in each generation, the fittest candidate solutions are able to reproduce with genetic modification. Figure \ref{fig:generation} illustrates this process, where the fitness of the candidate solutions is at first evaluated. Then, according to the concept of natural selection, the fittest candidate solutions are selected to the mating pool. The members of the mating pool create offspring, which are modified via genetic operators, such as mutation or genetic crossover. Finally, the offspring replace unfit candidate solutions in the population \cite{weatheritt2016}.

\begin{figure}[t]
    \centering
    \includegraphics{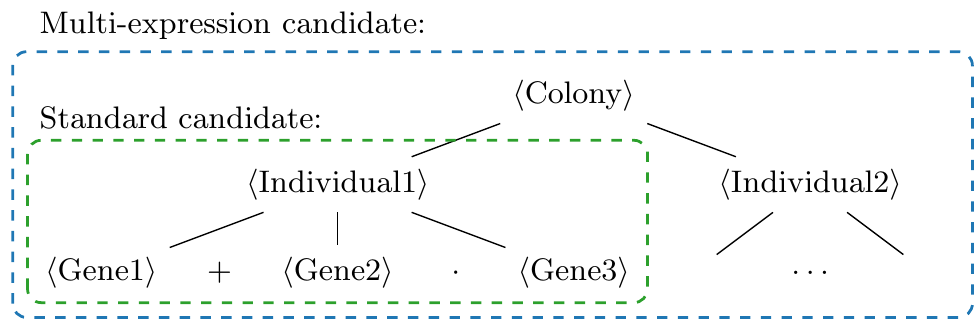}
    \caption{Exemplary candidate structure in standard (green) and extended (blue) EVE framework.}
	\label{fig:colony}
\end{figure}

\begin{figure}[t]
    \centering
    \includegraphics{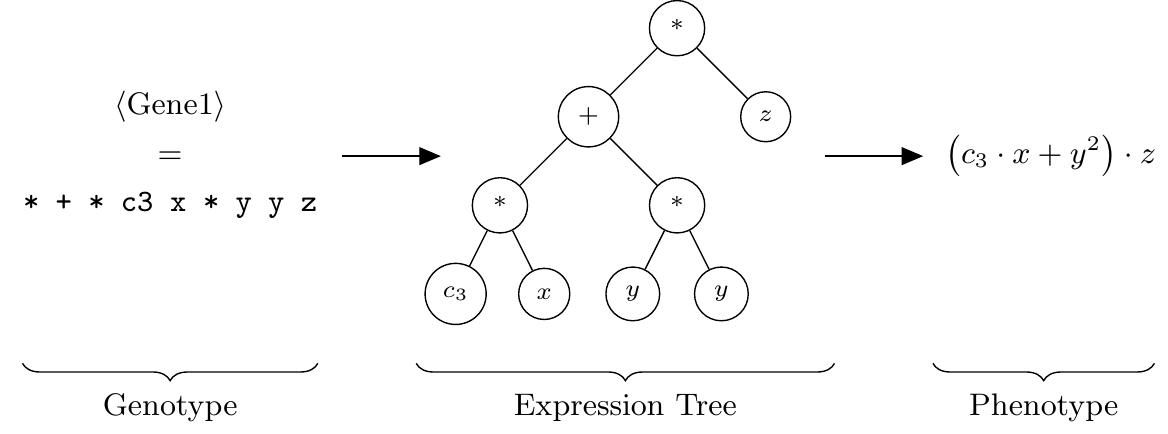}
    \caption{Exemplary translation from genotype to phenotype.}
	\label{fig:translation}
\end{figure}

In GEP, candidate solutions are referred to as individuals and each individual consists of multiple genes (see Figure \ref{fig:colony}, green box). Each gene is defined by its genotype, which is a linear string of symbols, such as problem-specific variables, mathematical operators or constants. The genotype of a gene can be translated to an expression tree, which represents an algebraic equation that is referred to as the gene's phenotype. Figure \ref{fig:translation} demonstrates the translation process for a gene consisting of the variables $x$, $y$ and $z$, addition and multiplication symbols and the constant $c_3$. In order to obtain the complete algebraic equation of a candidate solution, the phenotypes of its genes are linked. For details on the translation process, see \citet{weatheritt2016}.

In the original implementation of the EVE framework, the fitness of a candidate solution is evaluated internally, i.e. the fitness value is determined solely based on the candidate's algebraic equation and the high-fidelity training data. In the example of training an anisotropy model, calculating the mean squared error of the model prediction to the high-fidelity anisotropy data is considered an internal evaluation. In contrast, in an external evaluation, the fitness of a candidate depends on data from external software, which is executed during the training process. The CFD-driven training approach by \citet{zhao2020} is an external evaluation technique, as the candidate's algebraic equation is inserted in a RANS solver and is not evaluated explicitly. The fitness of the candidate solution is then calculated from the RANS results and the high-fidelity data.

\subsection{Multi-Expression Training}
\label{sec:multi-expression}

The goal of multi-expression training is to generate multiple closure models simultaneously in one training run. In combination with the CFD-driven training approach, the coupling and the coupled effect of the derived models on the RANS predictions are accounted for in the training process. This is a novel capability in data-driven turbulence modeling.

In order to train multiple models simultaneously, the concept of colonies is introduced. A colony consists of multiple individuals, where each individual represents one closure model (see Figure \ref{fig:colony}, blue box). Now a candidate solution is no longer one individual, but a group of individuals, namely one colony. Thus, a population of colonies is evolved in multi-expression training in contrast to a population of individuals, which was done previously. Two aspects are important to the concept of colonies:

\begin{itemize}
    \item All individuals of one colony share the same fitness value.
    \item Genetic material is only exchanged between individuals that represent the same closure model.
\end{itemize}

\noindent The first point results from the fact that multi-expression training is designed to take coupling effects of the trained models into account. Assigning separate fitness values to the individuals of one colony would neglect the impact of the coupling and is often not feasible in practical applications. Secondly, the exchange of genetic material during reproduction, e.g. via genetic crossover between two individuals of the mating pool (see Figure \ref{fig:generation}), is only allowed between individuals that represent the same closure model. This restriction is necessary to develop models of different dimensionality and with different independent variables. For example, a symbol that encodes the mean temperature gradient vector, which is important to train a turbulent heat flux model, might not be a valid component of a Reynolds stress anisotropy model. As a consequence, each individual of a colony is created from a unique set of variables, mathematical operators and constants.

\subsection{Multi-Objective Optimization}
\label{sec:multi-objective}

In the standard version of the EVE framework (see Section \ref{sec:standard}), only a single training objective can be specified. This translates to each candidate solution having one fitness value. In the case that multiple objectives are of interest, e.g. minimizing the mean squared error of a velocity profile and a turbulence kinetic energy profile in CFD-driven training, the fitness value is calculated as a weighted sum of the individual objectives. While the weighted sum approach is computationally efficient, there are two major disadvantages \citep{konak2006}:

\begin{itemize}
    \item The weights of the individual objectives need to be set before training.
    \item Candidate solutions that do not perform well on all objectives are likely discarded.
\end{itemize}

The weighting of multiple objective functions before running the actual model training is a non-trivial task, as the ideal ratios between the different training objectives that yield optimal models are difficult to predict. A poor selection of weights can limit the training outcome, as demonstrated later in Section \ref{sec:results}. Furthermore, candidate solutions that perform well on one objective but poorly on another objective have a low probability of mating, as the weighted sum of the objectives yields only an average fitness value. However, since individuals in GEP consist of multiple genes, replicating the genes that result in a good performance on one objective might benefit the training performance.

To overcome those limitations, a multi-objective optimization capability is added to the EVE framework. This capability allows candidate solutions to hold multiple fitness values, specifically one fitness value per training objective. To determine the candidates that are selected to the mating pool, an additional metric is required that balances the different fitness values. The Non-Dominated Sorting Genetic Algorithm II (NSGA-II) by \citet{deb2002} is utilized for this task and is introduced in the following.

NSGA-II is a genetic algorithm that is based on the concept of Pareto domination. One candidate solution (A) is said to dominate another candidate solution (B) if the following two conditions apply \citep{deb2011}:

\begin{enumerate}
    \item Solution (A) is not worse than solution (B) for all objective functions.
    \item Solution (A) is strictly better than solution (B) for at least one objective function.
\end{enumerate}

\begin{figure}
    \centering
    \includegraphics{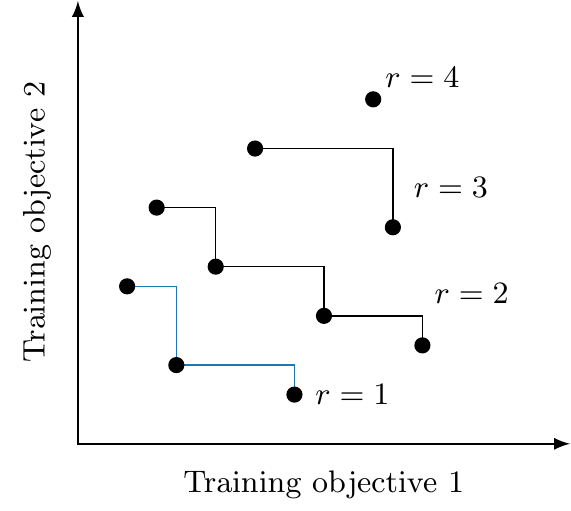}
    \caption{Exemplary ranking of candidate solutions for multi-objective optimization with highlighted Pareto front (blue).}
	\label{fig:pareto}
\end{figure}

\noindent When Pareto domination is applied to the entire population via pairwise comparison, each candidate solution is assigned a rank $r$. Within one rank, no solution dominates another solution. However, all solutions of rank $r = c_1$ dominate all solutions of rank $r = c_2$, where $c_1$ and $c_2$ are integer values and $c_1$ is smaller than $c_2$. The solutions of rank $r=1$ are not dominated by any other solution in the population and define what is commonly referred to as the Pareto front. Figure \ref{fig:pareto} illustrates an exemplary ranking of a population in the case of minimizing two objective functions. The implementation of NSGA-II in the EVE framework allows the specification of an arbitrary number of objective functions.

In addition to approximating the true Pareto front, i.e. the optimal solution to the specific optimization problem, the multi-objective optimization algorithm should aim to distribute solutions uniformly across the Pareto front and explore the extreme ends of the objective function space \citep{konak2006}. To achieve these aspects, the NSGA-II algorithm's crowding distance is introduced as a second metric. The crowding distance algorithm is applied to each Pareto rank separately and calculates the distance in objective function space of a solution to its nearest neighbors of the same rank \citep{deb2002}. By maximizing the crowding distances of the candidate solutions, a uniform spread across the Pareto front and the exploration of the extreme ends is incentivized.

For the implementation in the EVE framework, the crowding distance is inverted and normalized to a range between 0 and 1. A minimum value is now targeted for both this modified crowding distance and the Pareto rank, where a rank of $r=1$ indicates non-dominated solutions. Since Pareto ranks are integer values and modified crowding distances are positive values of order $\mathcal{O}(10^{-1})$, the two metrics are added to an extended Pareto rank $\widetilde{r}$. For example, a value of $\widetilde{r} = 3.05$ describes a candidate solution with Pareto rank $r = 3$ and a large distance to its nearest neighbors, while a value of $\widetilde{r} = 1.90$ is assigned to a solution in a crowded area of the Pareto front. Finally, the fitness of candidate solutions during natural selection is determined based on their extended Pareto rank. The increase in computational costs by introducing multi-objective optimization is small compared to the typical costs of RANS calculations in the CFD-driven training approach.

\section{Modeling Strategy and High-Fidelity Data}
\label{sec:strategy}

The novel capabilities of the EVE framework are demonstrated in this paper on two flow problems, namely a two-dimensional periodic hills (PH) flow and a one-dimensional vertical natural convection (VNC) flow. The PH case is selected because it is commonly used as a turbulence modeling test case. For this flow, we train one model for the Reynolds stress anisotropy and one model to correct the production term in the turbulence transport equations, as inspired by the work of \citet{schmelzer2019}, but in a coupled way. In order to demonstrate the capabilities of the presented concepts for a flow with a strong coupling between the momentum and thermal fields, one anisotropy model, one production correction model and one turbulent heat flux model are trained for the VNC case. The robustness of these models is assessed by testing the models at unseen flow conditions. All closure models are developed via multi-objective, multi-expression training using the CFD-driven training approach. The details on the flow cases, the closure models, the high-fidelity data for training and testing and the RANS solvers are presented in the following.

\subsection{Periodic Hills Flow}
\label{sec:periodic-hills}

\begin{figure}[!ht]
	\centering
    \resizebox{0.75\linewidth}{!}{
		\centering
        \includegraphics{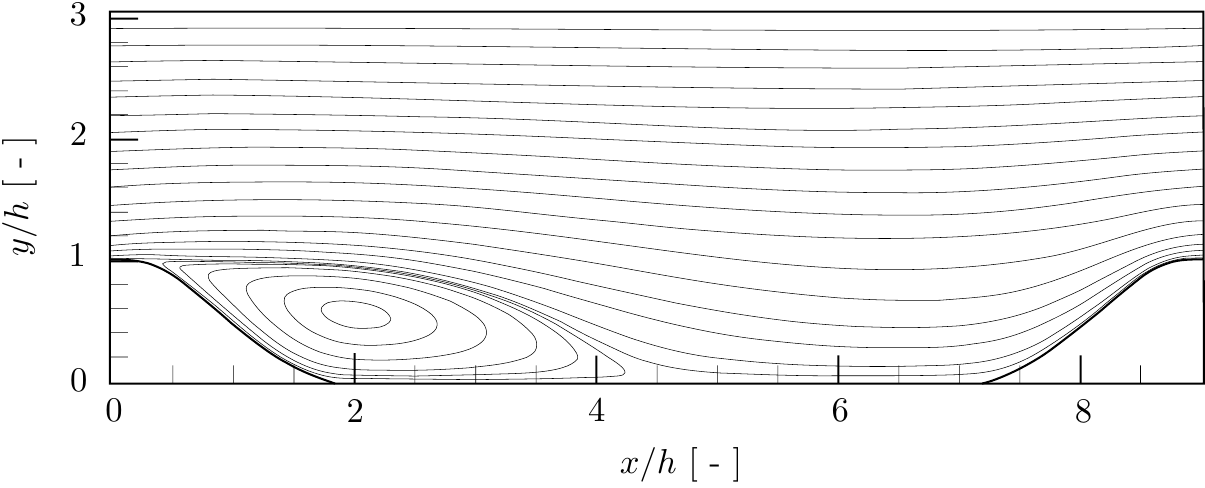}
	}
	\caption{Schematic of periodic hills domain and streamlines \citep{breuer2009}.}
	\label{fig:PH_domain}
\end{figure}

The two-dimensional periodic hills case at a Reynolds number of $Re = 10,595$ is a popular turbulence model benchmark case, which includes flow separation and reattachment (see Figure \ref{fig:PH_domain}). For an incompressible, constant density and zero gravity flow, the following RANS equations apply:

\begin{align}
    \frac{\partial\overline{u}_i}{\partial x_i} &= 0 \, , \\
    \overline{u}_j \frac{\partial\overline{u}_i}{\partial x_j} &= - \frac{\partial\overline{p}}{\partial x_i} + \frac{\partial}{\partial x_j} \left( \nu \frac{\partial\overline{u}_i}{\partial x_j} - \overline{u^\prime_i u^\prime_j} \right) \, ,
\end{align}

\noindent where $\overline{u}_i$ is the mean velocity, $\overline{p}$ is the density-corrected mean pressure and $\nu$ is the kinematic viscosity. The unclosed Reynolds stress term $\overline{u^\prime_i u^\prime_j}$ can be decomposed as

\begin{equation}
    \overline{u^\prime_i u^\prime_j} = \frac{2}{3} k \delta_{ij} + k a_{ij} \, ,
\end{equation}

\noindent where $a_{ij}$ is the anisotropy tensor, $k$ is the turbulence kinetic energy and $\delta_{ij}$ is the Kronecker delta. Most LEVMs assume a linear relation between $a_{ij}$ and the mean strain rate $S_{ij} = \frac{1}{2} \left(\frac{\partial\overline{u}_i}{\partial x_j} + \frac{\partial\overline{u}_j}{\partial x_i}\right)$ via

\begin{equation}
    a_{ij} = -2 \frac{\nu_t}{k} S_{ij} \, ,
    \label{eq:levm}
\end{equation}

\noindent where $\nu_t$ is the turbulent eddy viscosity. As a baseline turbulence model in Section \ref{sec:results}, we use the $k$-$\omega$ SST model by \citet{menter1994}, which solves the transport equations for $k$ and the specific dissipation rate $\omega$,

\begin{align}
    \frac{\partial k}{\partial t} + \overline{u}_j \frac{\partial k}{\partial x_j} &= P_k - \beta^* k \omega + \frac{\partial}{\partial x_j} \left( \left( \nu + \sigma_k \nu_t \right) \frac{\partial k}{\partial x_j} \right) \, , \label{eq:ktransport} \\
    \frac{\partial \omega}{\partial t} + \overline{u}_j \frac{\partial \omega}{\partial x_j} &= \frac{\gamma}{\nu_t} P_k - \beta \omega^2 + \frac{\partial}{\partial x_j} \left( \left( \nu + \sigma_\omega \nu_t \right) \frac{\partial \omega}{\partial x_j} \right) \nonumber \\
    &\qquad + 2 \left( 1 - F_1 \right) \sigma_{\omega 2} \frac{1}{\omega} \frac{\partial k}{\partial x_j} \frac{\partial \omega}{\partial x_j} \, , \label{eq:omegatransport}
\end{align}

\noindent in order to calculate $\nu_t$ via

\begin{equation}
    \nu_t = \frac{a_1 k}{\text{max}\left(a_1 \omega, S F_2\right)} \, .
    \label{eq:eddyviscosity}
\end{equation}

\noindent Here $S$ is the invariant measure of $S_{ij}$. The production of turbulence kinetic energy $P_k$ was limited by \citet{menter2003} to

\begin{equation}
    P_k = \text{min}\left(- k a_{ij} \frac{\partial\overline{u}_i}{\partial x_j}, 10 \beta^* \omega k \right) \, ,
    \label{eq:kproduction}
\end{equation}

\noindent while the remaining variables are applied according to their definition in \citet{menter1994}.

\subsubsection*{Modeling Strategy}

To address the inaccuracies of LEVMs mentioned in Section \ref{sec:introduction}, we develop a non-linear eddy viscosity model (NLEVM), which is formally equivalent to an explicit algebraic Reynolds stress model (EARSM), for the Reynolds stress anisotropy $a_{ij}$ in order to replace Eq.~\eqref{eq:levm}. The modeling approach is based on the integrity basis derived by \citet{pope1975} (see Eq.~\eqref{eq:integrity_basis}). For a two-dimensional flow, only three basis tensors are required to form a linear independent basis and there are two non-zero invariants:

\begin{align}
        V_{ij}^1 &= s_{ij} \, , &  I^1 &= s_{mn} s_{nm} \, , \nonumber \\
        V_{ij}^2 &= s_{ik} w_{kj} - w_{ik} s_{kj} \, , & I^2 &= w_{mn} w_{nm} \, , \label{eq:basistensors}\\
        V_{ij}^3 &= s_{ik} s_{kj} - \frac{1}{3} \delta_{ij} s_{mn} s_{nm} \, , \nonumber
\end{align}

\noindent where $s_{ij}$ and $w_{ij}$ are the normalized mean strain rate $S_{ij}$ and mean rotation rate $\Omega_{ij}$ tensors, as made dimensionless using the turbulent timescale $\tau = \frac{1}{\omega}$. The mean rotation rate tensor is defined by $\Omega_{ij} = \frac{1}{2} \left(\frac{\partial\overline{u}_i}{\partial x_j} - \frac{\partial\overline{u}_j}{\partial x_i}\right)$. In Section \ref{sec:results}, the EVE framework is utilized to discover the respective model form of the scalar functions $g^1$ to $g^3$ to model the anisotropy via

\begin{equation}
    a_{ij} = \sum\limits_{k=1}^{3} g^k\left(I^1, I^2\right)V_{ij}^k \, .
    \label{eq:anisotropyPH}
\end{equation}

Additionally, following \citet{schmelzer2019}, we train a second closure model to correct the production term $P_k$ in the turbulence transport equations Eq.~\eqref{eq:ktransport} and \eqref{eq:omegatransport}. Therefore, $P_k$ is extended in both equations to $\widetilde{P_k} = P_k + R$, where $R$ represents the local production correction. In accordance with the definition of $P_k$ in Eq.~\eqref{eq:kproduction}, the term $R$ is modeled as

\begin{equation}
    R = k a_{ij}^R \frac{\partial\overline{u}_i}{\partial x_j} \, .
    \label{eq:productioncorrection}
\end{equation}

\noindent The modeling approach for the tensor $a_{ij}^R$ is equivalent to the approach for the anisotropy $a_{ij}$:

\begin{equation}
    a^R_{ij} = \sum\limits_{k=1}^{3} h^k\left(I^1, I^2\right)V_{ij}^k \, ,
    \label{eq:productionPH}
\end{equation}

\noindent where the scalar functions $h^1$ to $h^3$ are derived by the EVE framework. 

To drive the development of the models for $a_{ij}$ and $R$ towards accurate predictions of the mean flow and turbulence quantities, two training objectives are defined:

\begin{align}
    J_1 &= \frac{1}{\overline{u}_b^{\,2}} \frac{1}{N_1} \sum\limits_{i=1}^{N_1} \left| \overline{u}_{1,i}^{\, \text{EVE}} - \overline{u}_{1,i}^{\, \text{HF}} \right|^2 \, , \\
    J_2 &= \frac{1}{\overline{u}_b^{\,4}} \frac{1}{N_1} \sum\limits_{i=1}^{N_1} \left| k_i^{\, \text{EVE}} - k_i^{\, \text{HF}} \right|^2 \, .
\end{align}

\noindent The first objective $J_1$ aims to minimize the mean squared error of the streamwise mean velocity $\overline{u}_{1}$ between the results based on the EVE models and the high-fidelity (HF) data over all training data points $N_1$. In a similar way, objective $J_2$ aims to minimize the mean squared error of the turbulence kinetic energy. The two objective functions are non-dimensionalized using the bulk velocity $\overline{u}_b$ at the hill crest.

\subsubsection*{High-Fidelity Data}

As training data for the model development, the highly resolved LES data of a PH flow at $Re = 10,595$ by \citet{breuer2009} is utilized. Figure \ref{fig:PH_domain} shows a schematic of the flow domain and the streamlines at the investigated Reynolds number. The number of training data points is $N_1=960$, which results from $N_{1,y}=160$ points in the wall-normal direction at six axial positions $x/h = \{0, 1, 2, 4, 6, 8\}$, where $h$ is the hill size. The number of points $N_{1,y}$ corresponds to the wall-normal dimension of the structured numerical grid of the RANS solver, which is of size $200\times160$. The open-source software OpenFOAM \cite{weller1998} is selected as a RANS solver to perform the external evaluations in the CFD-driven training. OpenFOAM is a finite-volume solver and its steady-state, incompressible algorithm is employed. All terms are discretized using a second-order central scheme, except for the divergence of the mean velocity, for which a first-order upwind scheme is applied. A thorough grid independence study was performed and the cell size at both walls is set to ensure a non-dimensional wall distance of $y^+ < 1$.

\subsection{Vertical Natural Convection Flow}
\label{sec:vertical-natural-convection}

\begin{figure}[!ht]
	\centering
    \resizebox{0.5\linewidth}{!}{
		\centering
        \includegraphics{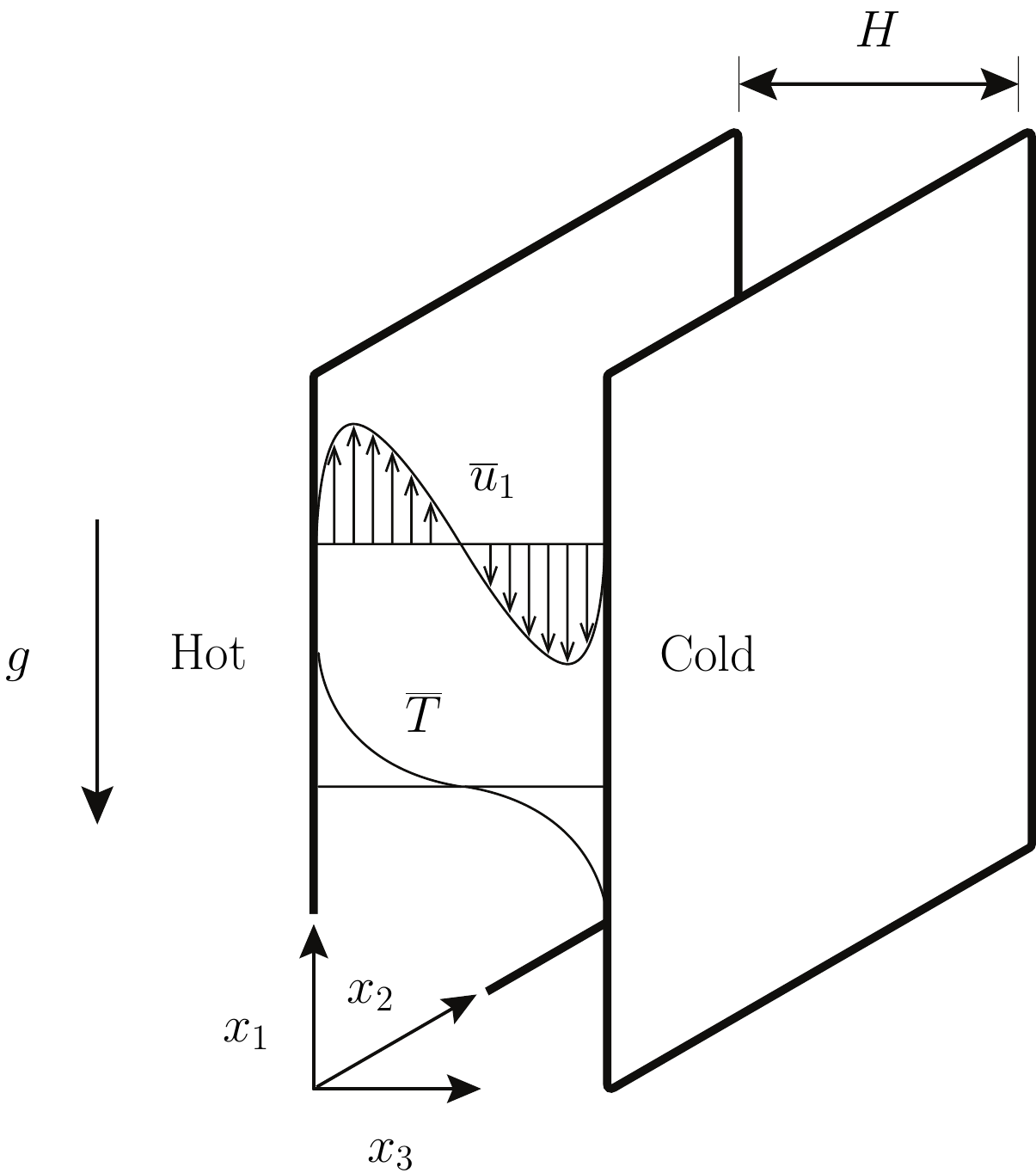}
	}    
	\caption{Schematic of vertical channel with differently heated walls and mean flow quantity plots \citep{ng2015}.}
	\label{fig:VNC_domain}
\end{figure}

The second flow problem, to which the novel capabilities of the EVE framework are applied, is natural convection flow in a vertical channel. A temperature difference between the channel walls causes a buoyancy-driven flow and the infinite extent of the walls in both streamwise and spanwise direction results a one-dimensional flow problem. The Rayleigh number, which describes the ratio of the viscosity to the buoyant forces, is set to $Ra=5.4\times10^5$ for training to ensure a turbulent bulk flow. A common approximation for this flow problem is the assumption of a constant density flow, where density variations are only considered via the thermal expansion coefficient $\beta$ \citep{versteegh1999}, which leads to the following RANS equations:

\begin{align}
    \frac{\partial\overline{u}_1}{\partial x_1} &= 0 \, , \\
    \frac{\partial}{\partial x_3} \left( \nu \frac{\partial\overline{u}_1}{\partial x_3} - \overline{u^\prime_1 u^\prime_3} \right) + g \beta \left(\overline{T} - T_0\right) &= 0 \, , \\
    \kappa \frac{\partial}{\partial x_3} \left(\frac{\partial\overline{T}}{\partial x_3} - \overline{u^\prime_3 T^\prime} \right) &= 0 \, .
\end{align}

\noindent The streamwise and wall-normal directions are indicated by the indices 1 and 3, respectively, $\overline{T}$ is the mean temperature, $T_0$ is a reference temperature at the channel center, $g$ is the gravitational acceleration and $\kappa$ is the thermal diffusivity. The Reynolds stress $\overline{u^\prime_1 u^\prime_3}$ and the turbulent heat flux $\overline{u^\prime_3 T^\prime}$ are modeled in order to close the RANS equations. Figure \ref{fig:VNC_domain} illustrates the flow domain with channel width $H$ and shows approximate plots of the mean flow quantities $\overline{u}_1$ and $\overline{T}$.

\subsubsection*{Modeling Strategy}

The modeling approach for $\overline{u^\prime_1 u^\prime_3}$ is, similar to the PH case, based on the anisotropy integrity basis derived by \citet{pope1975}. For the investigated, one-dimensional VNC flow, the Reynolds stress is equivalent to its anisotropic component. The basis tensor $V_{ij}^1$ is sufficient for a linear independent basis and the only non-zero invariant is $I_1$. Thus, the Reynolds stress is modeled as

\begin{equation}
    \overline{u^\prime_1 u^\prime_3} = k \: c\left(I^1\right) V_{13}^1 \, ,
\end{equation}

\noindent where $c\left(I^1\right)$ is the scalar function that is derived by the EVE framework. 

Additional similarities to the PH case are that the $k$-$\omega$ SST model is the baseline turbulence model and that a correction to its turbulence production term $P_k$ (see Eq.~\eqref{eq:productioncorrection}) is trained:

\begin{equation}
    R_T = k \: d\left(I^1\right) V_{13}^1 \: \frac{\partial\overline{u}_1}{\partial x_3} \, ,
\end{equation}

\noindent where the scalar function $d\left(I^1\right)$ is modeled during training.

The turbulent heat flux $\overline{u^\prime_3 T^\prime}$ is the third term that requires modeling. Similar to the integrity basis for the anisotropy tensor, an integrity basis for the turbulent scalar flux vector was presented in \citet{zheng1994}. The first vector of this basis is the mean temperature gradient non-dimensionalized by $\frac{\sqrt{k}}{\omega \Delta T}$:

\begin{equation}
    v^{1}_i = \frac{\sqrt{k}}{\omega \Delta T} \frac{\partial\overline{T}}{\partial x_i} \, ,
\end{equation}

\noindent where $\Delta T$ is the temperature difference between the channel walls. For a one-dimensional flow, the vector $v^1_i$ spans a linear independent basis in combination with the two invariants $I^1$ and $I_T^1$. The invariant $I^1$ is defined according to Eq.~\eqref{eq:basistensors} and $I_T^1$ is calculated from

\begin{equation}
    I_T^1 = (\frac{\sqrt{k}}{\omega \Delta T})^2 \frac{\partial\overline{T}}{\partial x_i} \frac{\partial\overline{T}}{\partial x_i} \, .
\end{equation}

\noindent Applying the modeling approach in the wall-normal direction yields the following:

\begin{equation}
    \overline{u^\prime_3 T^\prime} = \sqrt{k} \: e\left(I^1, I_T^1\right) v^1_3 \: \Delta T \, ,
\end{equation}

\noindent where $e\left(I^1, I_T^1\right)$ is again a scalar function that is modeled during training. By setting $e = - \frac{1}{Pr_t}$, the standard gradient diffusion hypothesis (SGDH) model is obtained \cite{leschziner2015}, which is applied as the baseline model to close the turbulent heat flux in Section \ref{sec:results}. The turbulent Prandtl number is assumed constant at $Pr_t = 0.9$. The objective functions

\begin{align}
    J_3 &= \frac{1}{\overline{u}_f^{\,2}} \frac{1}{N_2} \sum\limits_{i=1}^{N_2} \left| \overline{u}_{1,i}^{\, \text{EVE}} - \overline{u}_{1,i}^{\, \text{HF}} \right|^2 \, \label{eq:objective3} \\
    J_4 &= \frac{1}{\Delta T^{2}} \frac{1}{N_2} \sum\limits_{i=1}^{N_2} \left| \overline{T}_{i}^{\, \text{EVE}} - \overline{T}_{i}^{\, \text{HF}} \right|^2 \label{eq:objective4}
\end{align}

\noindent are defined to minimize the dimensionless mean squared error of the mean streamwise velocity $\overline{u}_{1}$ and the mean temperature $\overline{T}$ between the EVE model results and the high-fidelity data over all training data points $N_2$, respectively. The free fall velocity $\overline{u}_f$ is calculated from $\overline{u}_f = \sqrt{g \beta \Delta T H}$.

\subsubsection*{High-Fidelity Data}

The DNS data for the introduced VNC flow was created by \citet{ng2015} across four orders of Rayleigh number magnitude ($10^5$ to $10^9$). We utilize the data at $Ra = 5.4 \times 10^5$ for training and put the data at $Ra = \{1 \times 10^5, 2 \times 10^6, 5 \times 10^6, 2 \times 10^7, 1 \times 10^8\}$ aside for testing the developed models. The number of training data points is $N_2=201$, which corresponds to the numerical grid size of the RANS solver in the wall-normal direction. The RANS solver that is applied to perform the external evaluations for the VNC flow is a steady-state in-house code, which uses second-order central discretization schemes and was extensively tested for application in industry projects \cite{xu2021}.

\section{Results}
\label{sec:results}

In the following, the results of developing multiple closure models for the PH case at $Re = 10,595$ and the VNC case at $Ra = 5.4 \times 10^5$, as presented in Section \ref{sec:strategy}, are discussed. The models are trained using the novel multi-expression training and multi-objective optimization capabilities of the EVE framework, which were introduced in Section \ref{sec:methodology}. An overview of the training runs is presented in Table \ref{tab:training_runs}. For both training runs, a population of 100 colonies is evolved over 250 generations to minimize the respective training objectives. Each colony consists of one individual per trained expression and 4 genes per individual. In order to develop easily implementable and interpretable expressions, the symbols available to build the expressions are limited to addition ($+$), subtraction ($-$) and multiplication ($\times$) symbols, the constant coefficients $-1$, $1$ and $2$ and five random coefficients in the range from $-1$ to $1$ (in addition to the respective input symbols, e.g. $I^1$ and $I^2$ to model the scalar functions $g^k$).

Based on this configuration, the two multi-objective training runs perform around 10,000 RANS evaluations each, which results in 4800 and 53 core hours of computational costs for the PH and VNC cases, respectively. While multi-expression training does not increase the computational costs compared to training the models individually, multi-objective optimization changes the costs, relative to a single-objective training run, by a factor of 1.7 for the PH case and a factor of 2.6 for the VNC case. Due to the high computational costs of the training runs, the reported hyperparameters are selected based on the authors' experience from previous studies \cite{weatheritt2016,sandberg2018,zhao2020}.

\begin{table}[h]
	\centering
	\caption{Overview of multi-objective EVE training runs}
	\begin{tabular}{l l l l}
        \toprule
        Flow case & Periodic hills & Vertical natural convection \\
        \midrule
        Models/expressions & $a_{ij}$, $R$ &  $\overline{u^\prime_1 u^\prime_3}$, $R_T$, $\overline{u^\prime_3 T^\prime}$ \\
        Training objectives & $J_1(\overline{u}_1)$, $J_2(k)$ &  $J_3(\overline{u}_1)$, $J_4(\overline{T})$ \\
        High-fidelity data & LES, \citet{breuer2009} & DNS, \citet{ng2015} \\
        Training parameter & $Re = 10,595$ & $Ra = 5.4 \times 10^5$ \\
        \bottomrule
	\end{tabular}
	\label{tab:training_runs}
\end{table}

\subsection{Periodic Hills Flow}
\label{sec:PHresults}

The extended EVE framework is applied to derive algebraic equations for the anisotropy tensor $a_{ij}$ and the turbulence production correction $R$, while taking the coupled effects of the two models into consideration and benefiting from the multi-objective optimization capability. The resulting models are referred to as EVE-MO models. As points of reference, the models trained by \citet{schmelzer2019} on the PH case (SCH), which are considered the state-of-the-art two-equation algebraic turbulence model for this case, and the models derived by the EVE framework using single-objective optimization (EVE-SO) are included in the following analysis. The SCH solution is a set of models that also consist of one model for $a_{ij}$ and one model for $R$, but the models' coupled effects were not accounted for and CFD calculations were only performed during the final model selection from a candidate pool, but not during the training process. The only difference of the EVE-SO training compared to the EVE-MO training is that the objective functions were added up in a weighted sum approach and single-objective optimization was performed.

\begin{figure}[ht]
    \centering
    \includegraphics[scale=0.75]{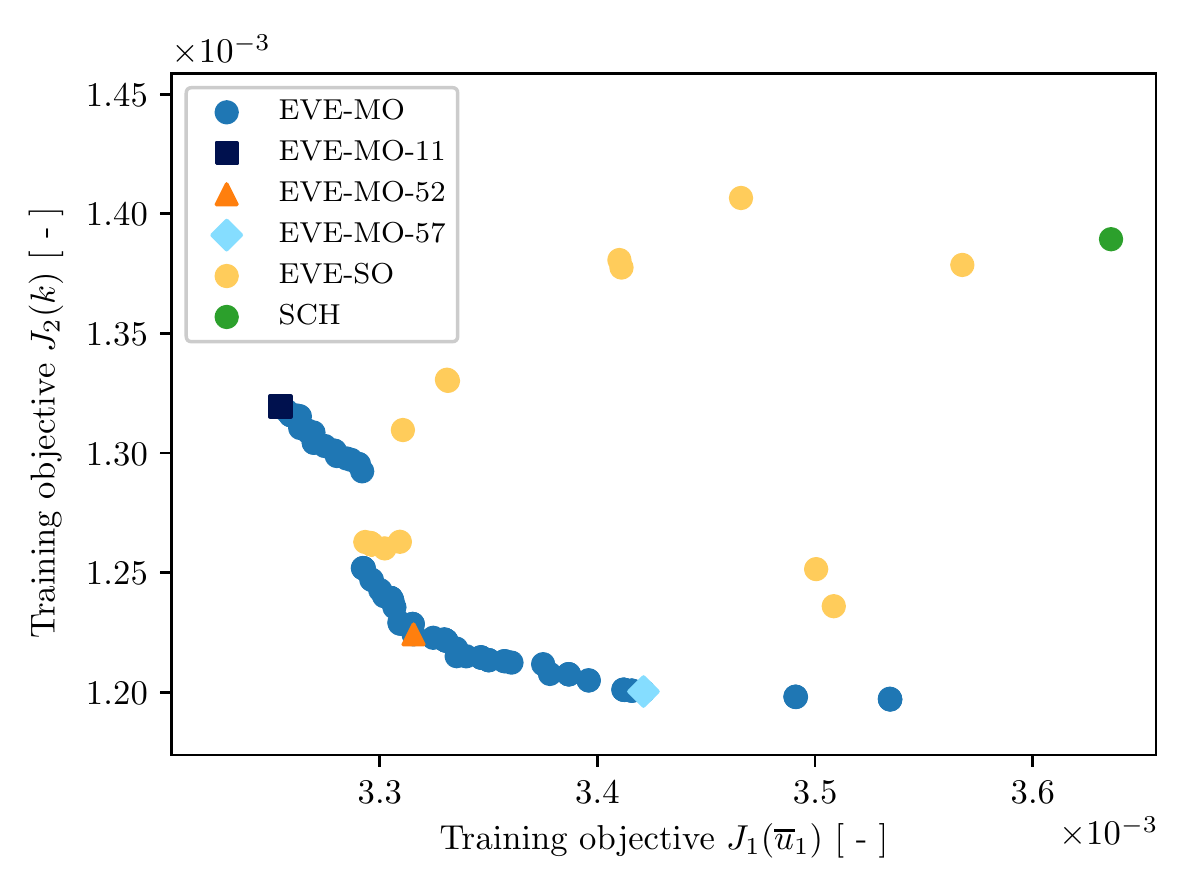}
    \caption{Performance of multi-objective and single-objective EVE models and models by \citet{schmelzer2019} on training objectives $J_1$ and $J_2$ for PH case.}
    \label{fig:PH_pareto}
\end{figure}

Figure \ref{fig:PH_pareto} shows the performance of the models resulting from these training strategies evaluated on the objective functions $J_1$ and $J_2$. The Pareto front of the EVE-MO training, i.e. the candidate solutions with rank $r=1$, is displayed in blue. The distribution of candidate solutions across the Pareto front is nearly uniform and the extreme ends are explored, e.g. by two candidate solutions around $J_1 = 3.5 \cdot \SI{e-3}{}$, which is indicative for the convergence of the multi-objective optimization algorithm. In contrast, the EVE-SO training, for which the 25 fittest candidate solutions are plotted in yellow, focuses on a narrow region of the Pareto front, which is influenced by the weights in the weighted sum approach. Thus, the choice for selecting different promising candidate solutions based on user preferences is very limited. From the EVE-MO training results, we can select specific candidate solutions that focus on reducing the mean velocity error (EVE-MO-11, dark blue), reducing the turbulence kinetic energy error (EVE-MO-57, light blue) or identifying the best compromise between the two training objectives (EVE-MO-52, orange). The specific algebraic equations for $a_{ij}$ and $R$ of the selected candidates are listed in Tables \ref{tab:PH_aij_expressions} and \ref{tab:PH_R_expressions} of \ref{sec:appendix}, respectively.

Both EVE training strategies result in models that improve over the performance of the set of SCH models, which is plotted in green. Table \ref{tab:PH_comparison} lists the relative difference in the training objectives $J_1$ and $J_2$ of the three selected EVE-MO solutions as compared to the baseline $k$-$\omega$ SST model (BSL) and the SCH solution. A reduction in mean streamwise velocity error of more than $55 \si{\percent}$ and in turbulence kinetic energy error of more than $65 \si{\percent}$ over the BSL model is achieved by all three EVE-MO solutions. Compared to the set of SCH models, a reduction in $\overline{u}_1$ prediction error by up to $10.5 \si{\percent}$ and in $k$ prediction error by up to $13.6 \si{\percent}$ is possible. However, according to the concept of Pareto dominance, an improvement of one training objective leads to a performance reduction on the other objective, as all three candidates have the same Pareto rank. The candidate EVE-MO-52 achieves the largest combined error reduction of $9.7 \si{\percent}$ compared to the state-of-the-art SCH solution.

\begin{table}[h]
	\centering
	\caption{Performance comparison of EVE-MO solutions for PH case}
	\begin{tabular}{l l l l l}
        \toprule
         & $\Delta J_1^{\text{BSL}}(\overline{u}_1)$ & $\Delta J_2^{\text{BSL}}(k)$ & $\Delta J_1^{\text{SCH}}(\overline{u}_1)$ & $\Delta J_2^{\text{SCH}}(k)$ \\
        \midrule
        EVE-MO-11 & $-58.09 \si{\percent}$ & $-65.49 \si{\percent}$ & $-10.49 \si{\percent}$ & $-5.03 \si{\percent}$ \\
        EVE-MO-52 & $-57.28 \si{\percent}$ & $-67.98 \si{\percent}$ & $-8.81 \si{\percent}$ & $-11.89 \si{\percent}$ \\
        EVE-MO-57 & $-55.94 \si{\percent}$ & $-68.61 \si{\percent}$ & $-5.91 \si{\percent}$ & $-13.61 \si{\percent}$ \\
        \bottomrule
	\end{tabular}
	\label{tab:PH_comparison}
\end{table}

\begin{figure}[ht]
    \centering
    \includegraphics{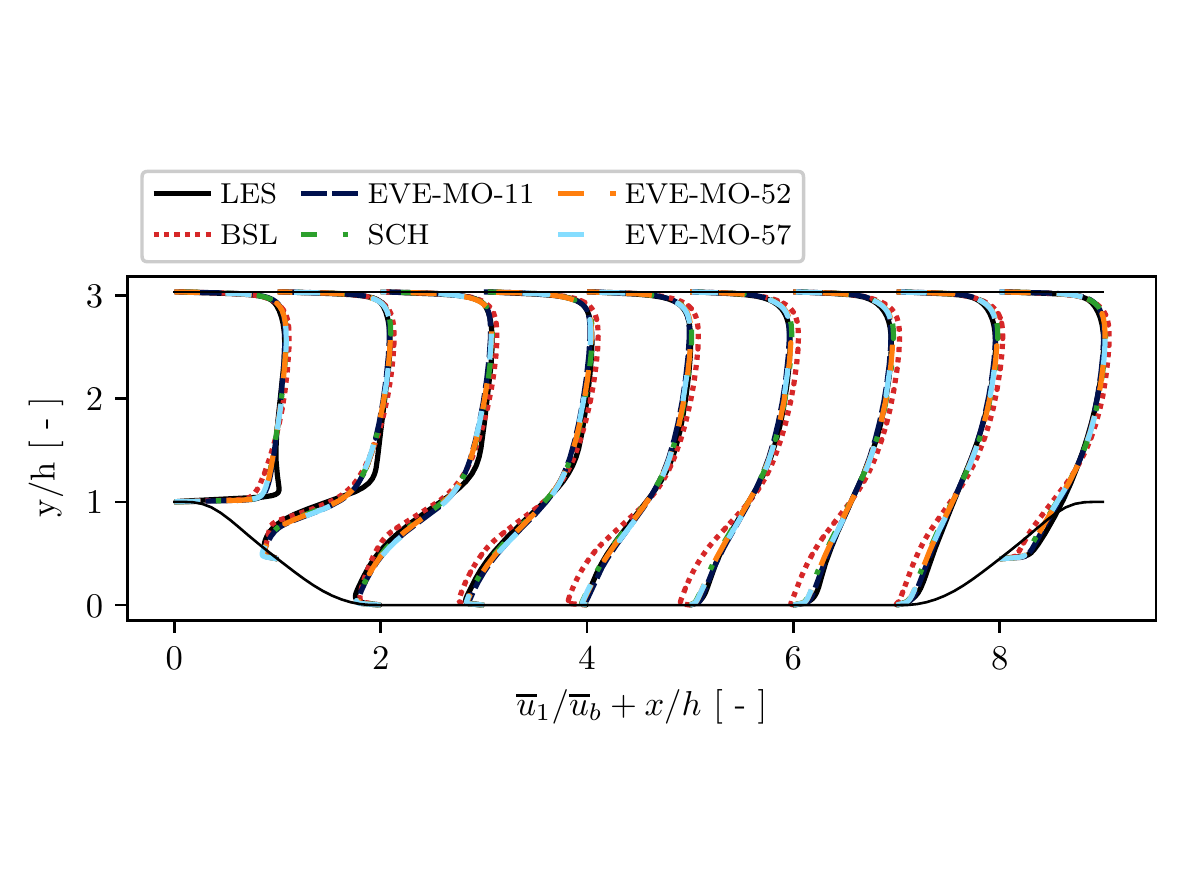}
    \vspace{-2cm}
    \caption{Wall-normal profiles of mean streamwise velocity in PH domain.}
    \label{fig:PH_profiles_u}
\end{figure}

\begin{figure}[ht]
    \centering
    \includegraphics{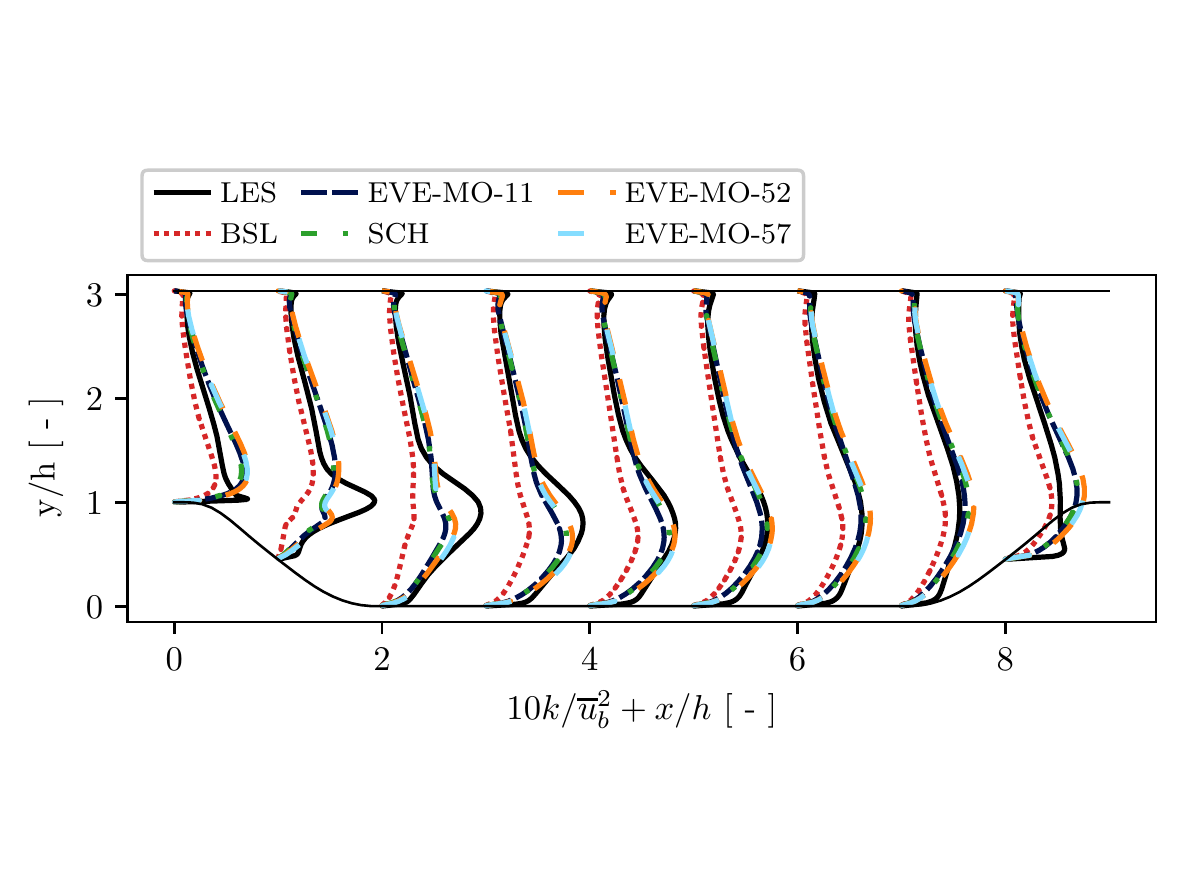}
    \vspace{-2cm}
    \caption{Wall-normal profiles of turbulence kinetic energy in PH domain.}
    \label{fig:PH_profiles_k}
\end{figure}

An overview of the $\overline{u}_1$ and $k$ profiles predicted by the selected EVE-MO models throughout the PH domain is presented in Figures \ref{fig:PH_profiles_u} and \ref{fig:PH_profiles_k}, respectively. The predictions of the BSL and SCH models as well as the LES training data are added for reference and all profiles are non-dimensionalized using the bulk velocity $\overline{u}_b$. All three EVE-MO models show mean velocity predictions that are clearly an improvement over the BSL model, but are very similar to the SCH solution. The shortcomings of the set of SCH models in the $\overline{u}_1$ profile predictions, e.g. around $y/h = 1$ at the axial positions $x/h = \{0, 1, 2\}$, are also not corrected under the current approach.

In order to locate the improvements shown in Table \ref{tab:PH_comparison}, the differences in mean squared $\overline{u}_1$ and $k$ prediction errors of the individual EVE-MO models compared to the SCH solution are summed up in the streamwise direction:

\begin{align}
    \sum \Delta J_1^{SCH} &= \frac{1}{\overline{u}_b^{\,2}} \sum\limits_{\frac{x}{h}=0}^{8} \left| \overline{u}_{1,i}^{\, \text{EVE}} - \overline{u}_{1,i}^{\, \text{HF}} \right|^2 - \left| \overline{u}_{1,i}^{\, \text{SCH}} - \overline{u}_{1,i}^{\, \text{HF}} \right|^2 \, , \\
    \sum \Delta J_2^{SCH} &= \frac{1}{\overline{u}_b^{\,4}} \sum\limits_{\frac{x}{h}=0}^{8} \left| k_{i}^{\, \text{EVE}} - k_{i}^{\, \text{HF}} \right|^2 - \left| k_{i}^{\, \text{SCH}} - k_{i}^{\, \text{HF}} \right|^2 \, .
\end{align}

\begin{figure}[ht]
    \centering
    \includegraphics[scale=0.75]{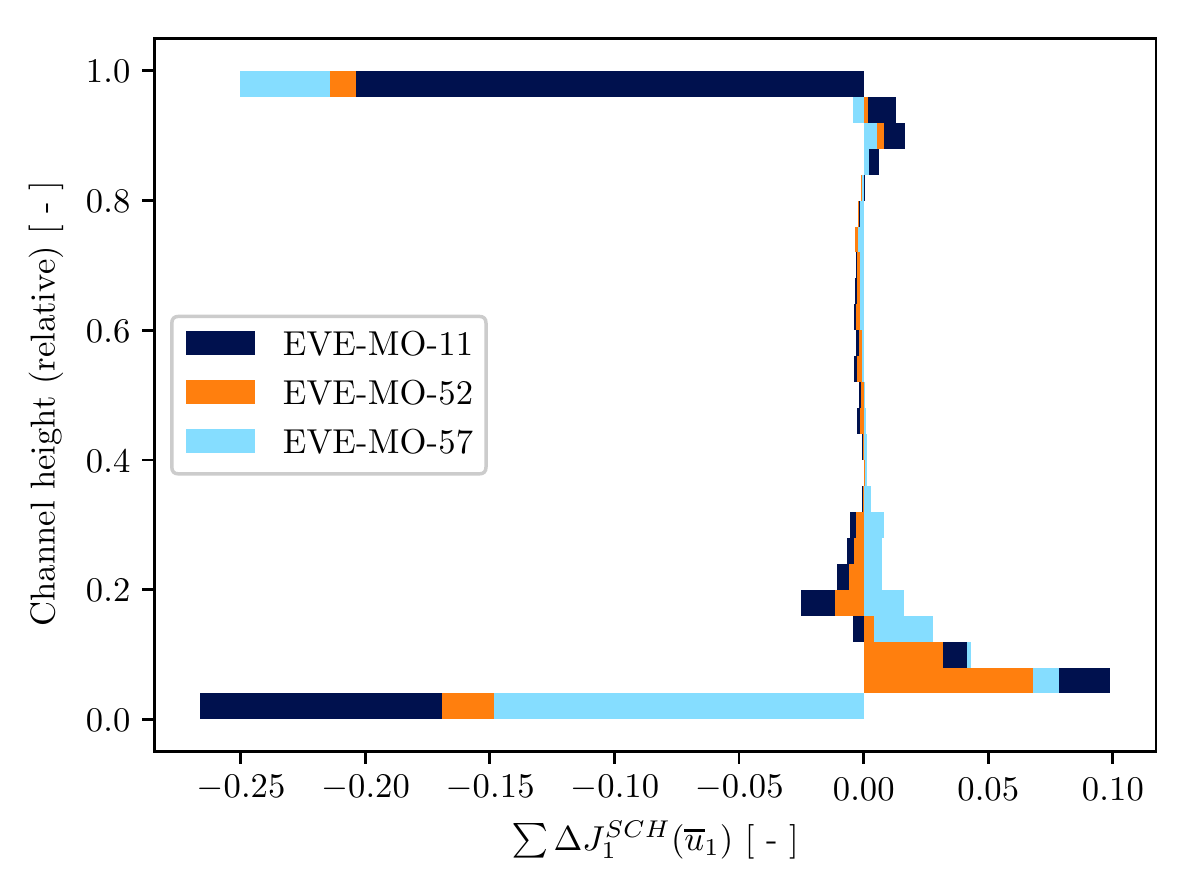}
    \caption{Histogram of mean squared $\overline{u}_1$ error difference to SCH predictions for PH case.}
    \label{fig:PH_hist_u}
\end{figure}

\noindent Figure \ref{fig:PH_hist_u} shows a histogram of $\sum \Delta J_1^{SCH}$ based on the relative channel height, revealing that the largest improvements in $\overline{u}_1$ prediction occur very close to the channel walls. From a relative channel height of $4 \si{\percent}$ up to around $30 \si{\percent}$, which is roughly the size of the recirculation bubble, the cumulated $\overline{u}_1$ prediction accuracy is decreased compared to the set of SCH models, especially for the high-$J_1$ EVE-MO-57 models.

\begin{figure}[ht]
    \centering
    \includegraphics[scale=0.75]{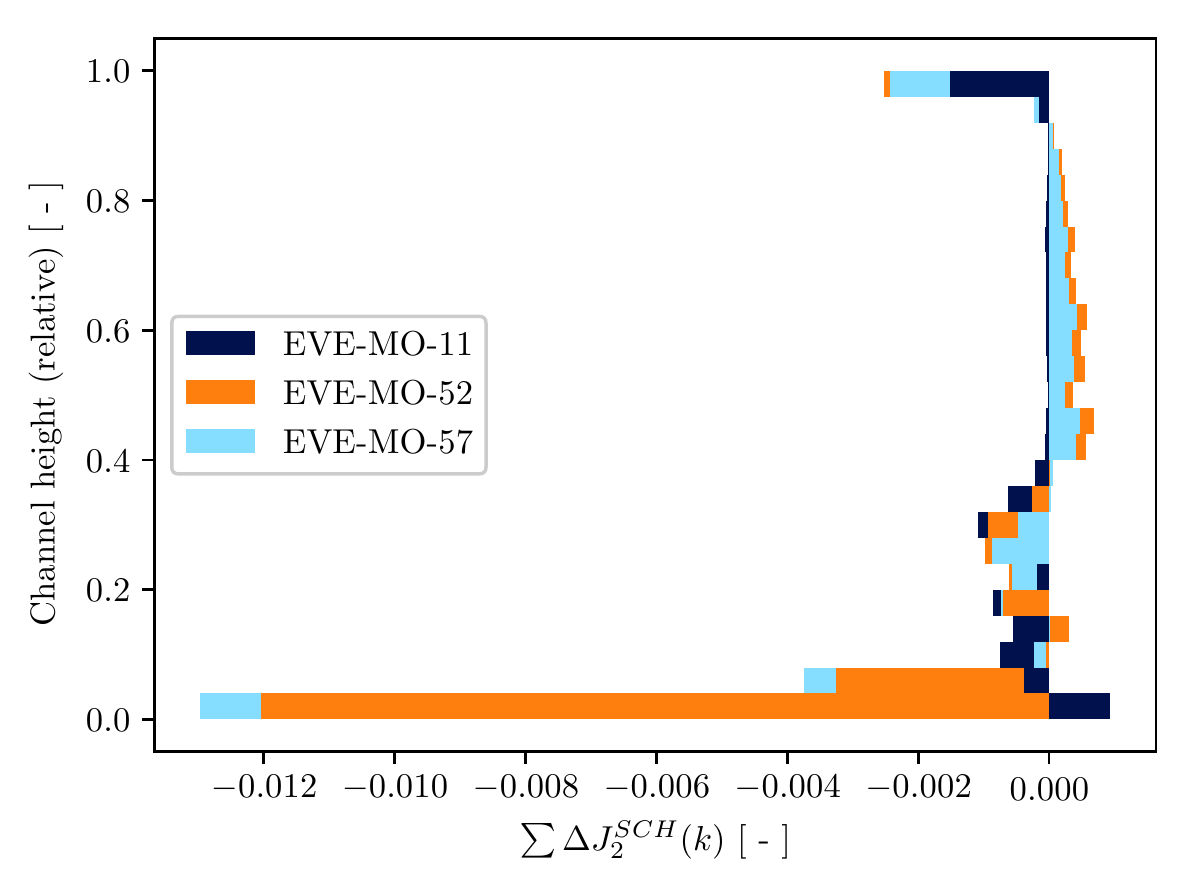}
    \caption{Histogram of mean squared $k$ error difference to SCH predictions for PH case.}
    \label{fig:PH_hist_k}
\end{figure}

Consistently, the $k$ profiles in Figure \ref{fig:PH_profiles_k} show a significant improvement over the BSL model for all three EVE-MO models. Furthermore, the strong peaks of $k$ from $x/h = 1$ to $x/h = 4$ are predicted slightly more accurately by the EVE-MO-52 and EVE-MO-57 models compared to the set of SCH models. Figure \ref{fig:PH_hist_k} confirms this improvement up to a channel height of $40 \si{\percent}$. The largest improvements in $k$ predictions are, however, again attained close to the channel walls.

In summary, the models derived with the novel EVE framework extensions are able to improve by near $10 \si{\percent}$ over the state-of-the-art SCH solution, which already shows very accurate $\overline{u}_1$ and reasonable $k$ predictions. Important areas of $\overline{u}_1$ inaccuracies were, however, not improved despite more complex expressions for $a_{ij}$ and $R$ (see Tables \ref{tab:PH_aij_expressions} and \ref{tab:PH_R_expressions}). Additionally, Figure \ref{fig:PH_profiles_k} illustrates that the $k$ profiles were scaled to achieve a better fit of the LES training data, but the desired improvements in the profile shape, e.g. at $x/h = 1$, were not attained. Due to the qualitatively similar predictions of the selected EVE-MO models compared to the set of SCH models, which was tested successfully on two different flow problems \cite{schmelzer2019}, no additional testing of the robustness of these models is performed.

In the view of the authors, the coupled effects of the $a_{ij}$ and $R$ models on the mean flow quantities are not very strong. Thus, the potential for improvement by the multi-expression training is limited. This argument is supported by the work of \citet{schmelzer2019}, who derived spatial fields for $a_{ij}$ and $R$ in a decoupled way. Inserting those fields in a CFD solver resulted in excellent $\overline{u}_1$ predictions. While spatial fields have a limited usefulness, as an application to different geometries is not feasible, those results illustrate that excellent predictions can be achieved for this particular flow problem without considering the coupling effects of the closure models.

\begin{figure}[ht]
    \centering
    \includegraphics[scale=0.75]{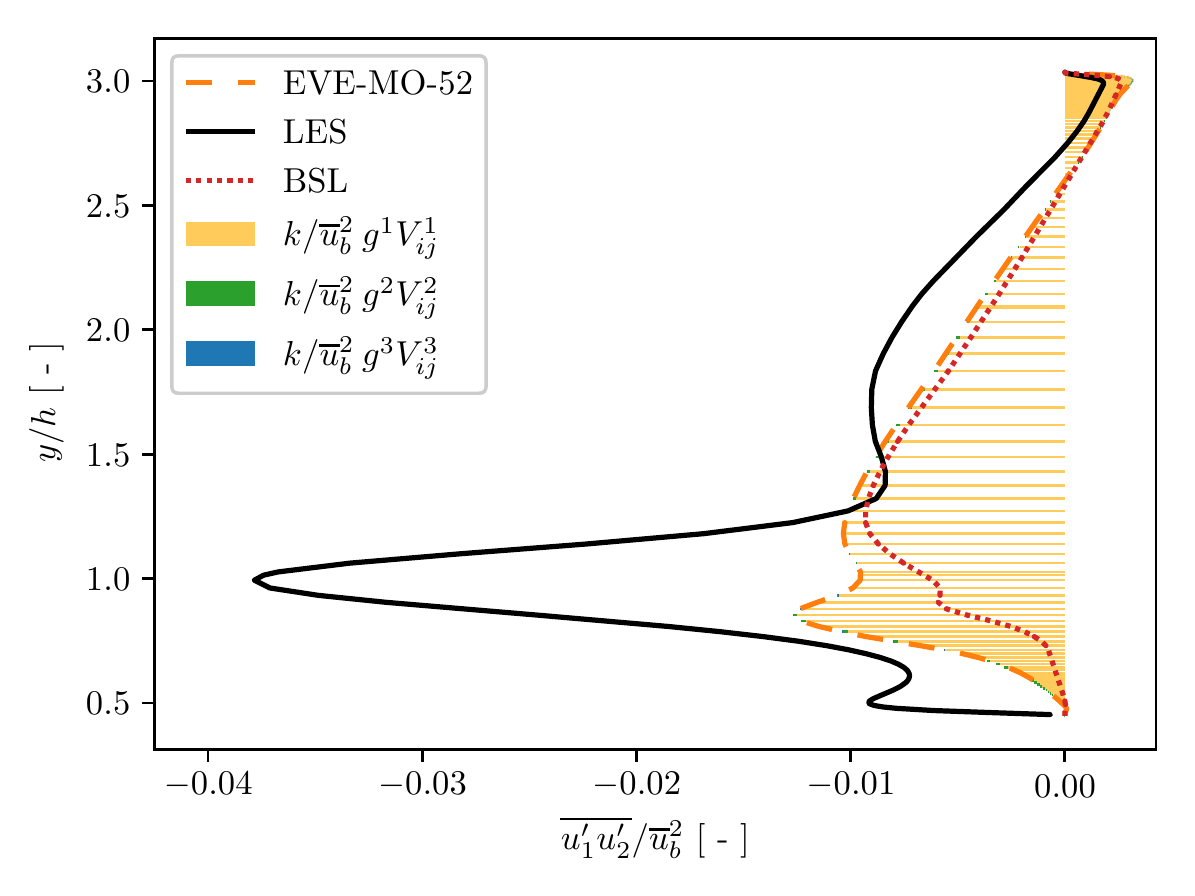}
    \caption{Wall-normal profiles of Reynolds shear stress and basis tensor contributions to EVE-MO-52 anisotropy model at $x/h = 1$ in PH domain.}
    \label{fig:PH_profiles_rs12}
\end{figure}

The remaining inaccuracies in predicting the $\overline{u}_1$ profiles are likely connected to underpredictions of the Reynolds shear stress $\overline{u^\prime_1 u^\prime_2}$. For the BSL model, weak turbulent mixing in the separated shear layer due to underpredicted $\overline{u^\prime_1 u^\prime_2}$ values was identified to cause inaccurate mean velocity predictions \cite{jakirlic2001}. Figure \ref{fig:PH_profiles_rs12} demonstrates that the EVE-MO-52 predictions of $\overline{u^\prime_1 u^\prime_2}$ at $x/h = 1$ show improvement compared to the BSL model. Relative to the LES data, however, a large underprediction persists. Additionally, the contributions of the individual basis tensors $V_{ij}^k$, multiplied by the derived scalar functions $g^k$ of the EVE-MO-52 anisotropy model and the turbulence kinetic energy $k$, are plotted and show the linear tensor $V_{ij}^1$ as the dominant contributor. Minimal contributions come from the non-linear tensors $V_{ij}^2$ and $V_{ij}^3$. This shortage of non-linear tensor contribution could explain why the $\overline{u}_1$ and $k$ profile shapes are not significantly improved compared to the linear BSL model at $x/h = 1$. This aspect further suggests that the features in Eq.~\eqref{eq:basistensors} are not sufficient to represent the mentioned spatial fields accurately. To further improve turbulence models for the PH case, a selection of additional input features seems necessary, for which \citet{wu2018} suggest a systematic approach. This objective is, however, beyond the scope of the current paper, as new features could be beneficial for both single-expression and multi-expression training approaches.

\subsection{Vertical Natural Convection Flow}
\label{sec:VNCresults}

The second training case for the extended EVE framework is the VNC flow at $Ra = 5.4 \times 10^5$ described in Section \ref{sec:vertical-natural-convection}, for which stronger coupled effects of the trained closure models are expected relative to the PH case. The multi-expression training and multi-objective optimization capabilities described in Section \ref{sec:methodology} are applied to train expressions for $\overline{u^\prime_1 u^\prime_3}$, $R_T$ and $\overline{u^\prime_3 T^\prime}$. The training objectives $J_3$ and $J_4$ are minimized to fit the predicted $\overline{u}_1$ and $\overline{T}$ profiles to the DNS training data. In the following analysis, the extended EVE framework (EVE-MO) results are compared to the results of the baseline $k$-$\omega$ SST and SGDH models (BSL), a single-objective EVE training approach (EVE-SO) and a single-objective, single-expression EVE training approach (EVE-SO-SE). As described previously, the EVE-SO training combines the training objectives in a weighted sum approach and is otherwise unchanged compared to the EVE-MO training. In contrast, the EVE-SO-SE approach uses the combined training objective for single-objective optimization and derives the closure models independently from each other via CFD-driven training. This approach represents the current state-of-the-art training strategy.

\begin{figure}[ht]
    \centering
    \includegraphics[scale=0.75]{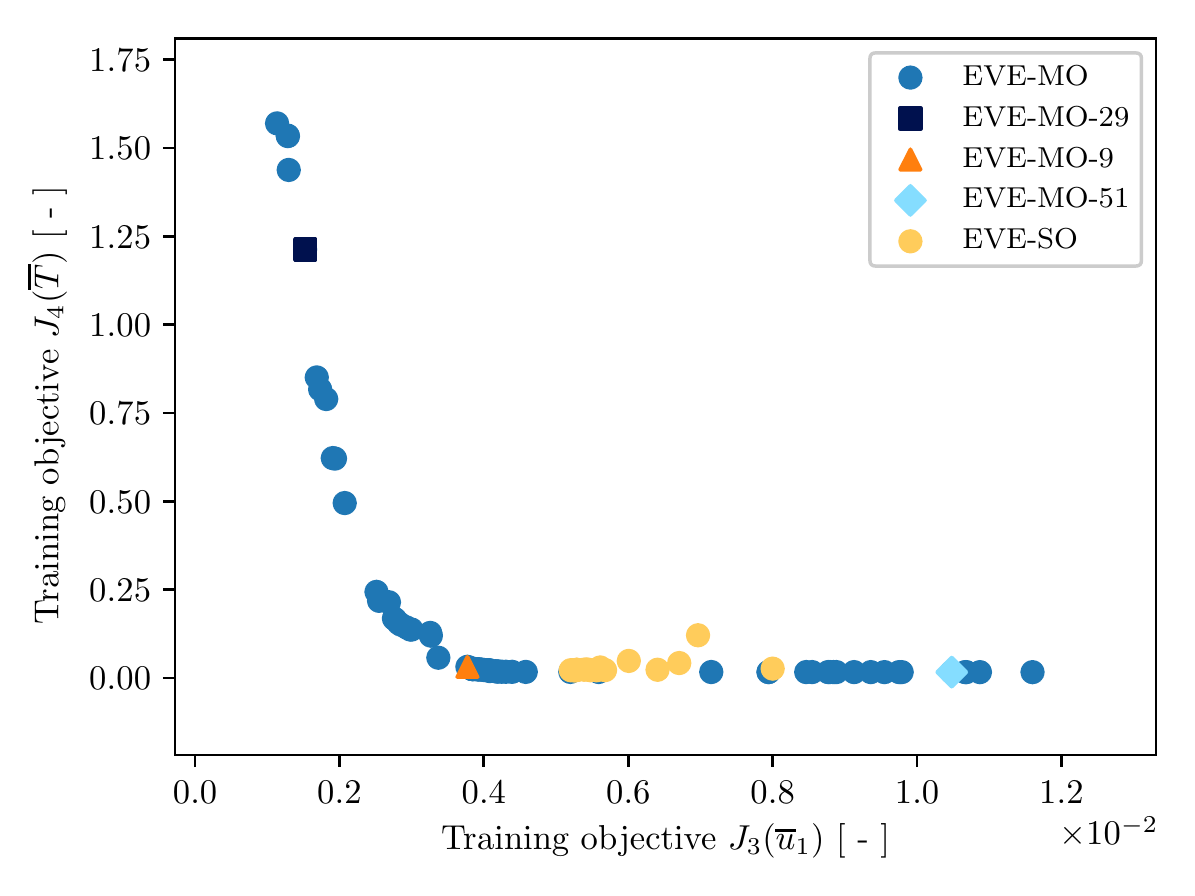}
    \caption{Performance of multi-objective and single-objective EVE models on training objectives $J_3$ and $J_4$ for VNC case at $Ra = 5.4 \times 10^5$.}
    \label{fig:VNC_pareto}
\end{figure}

Figure \ref{fig:VNC_pareto} shows the performance of the resulting EVE-MO and EVE-SO models with respect to the training objectives $J_3$ and $J_4$ at $Ra = 5.4 \times 10^5$. The corresponding values of the EVE-SO-SE and BSL models are located outside the displayed parameter ranges. The Pareto front of the EVE-MO training is plotted in blue and follows a uniform distribution reasonably well. In contrast, the 25 fittest candidate solutions of the EVE-SO training, which are displayed in yellow, cover only a narrow region of the Pareto front. The candidate solutions EVE-MO-29 (dark blue), EVE-MO-51 (light blue) and EVE-MO-9 (orange) are selected based on their error reduction of the mean velocity $\overline{u}_1$, the mean temperature $\overline{T}$ and the combined training objective, respectively. 

\begin{figure}[ht]
    \centering
    \includegraphics[scale=0.75]{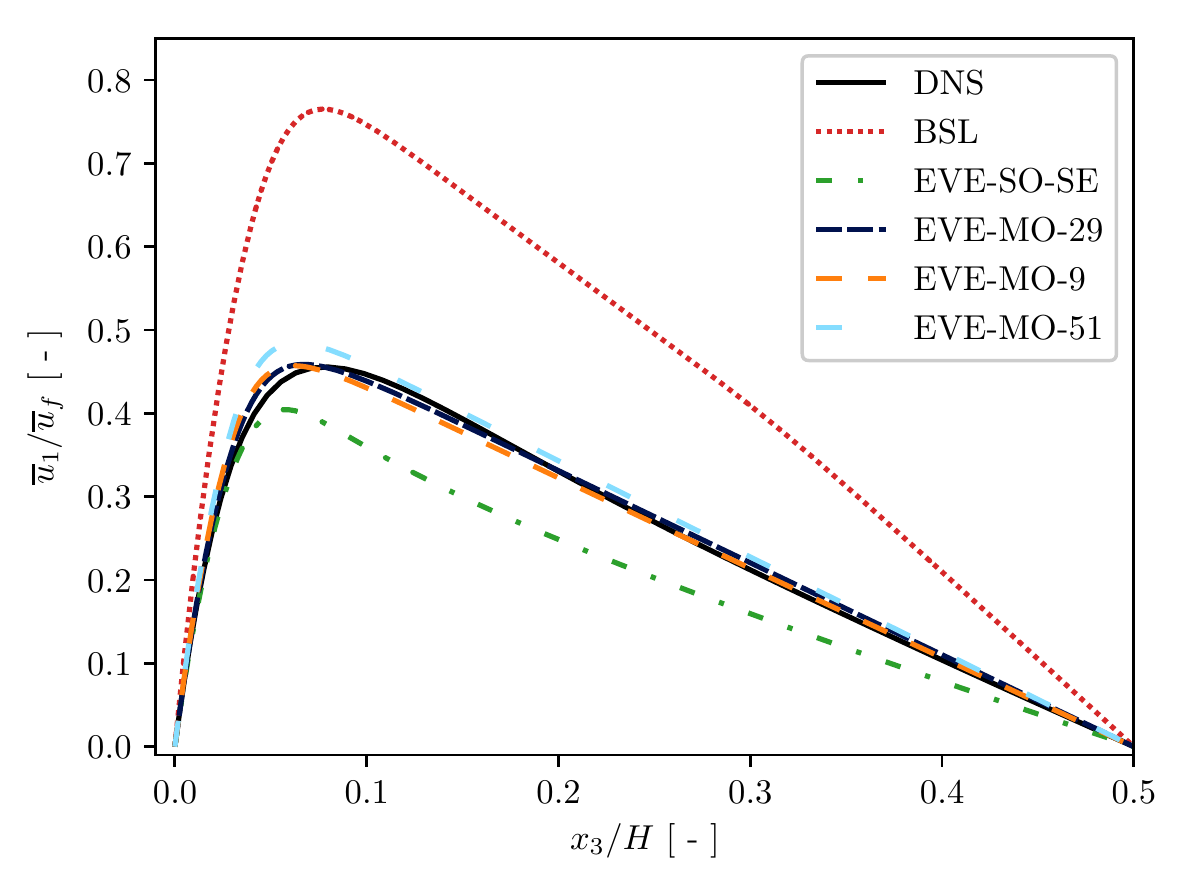}
    \caption{Wall-normal profiles of mean streamwise velocity for VNC case at $Ra = 5.4 \times 10^5$.}
    \label{fig:VNC_profiles_u}
\end{figure}

\begin{figure}[ht]
    \centering
    \includegraphics[scale=0.75]{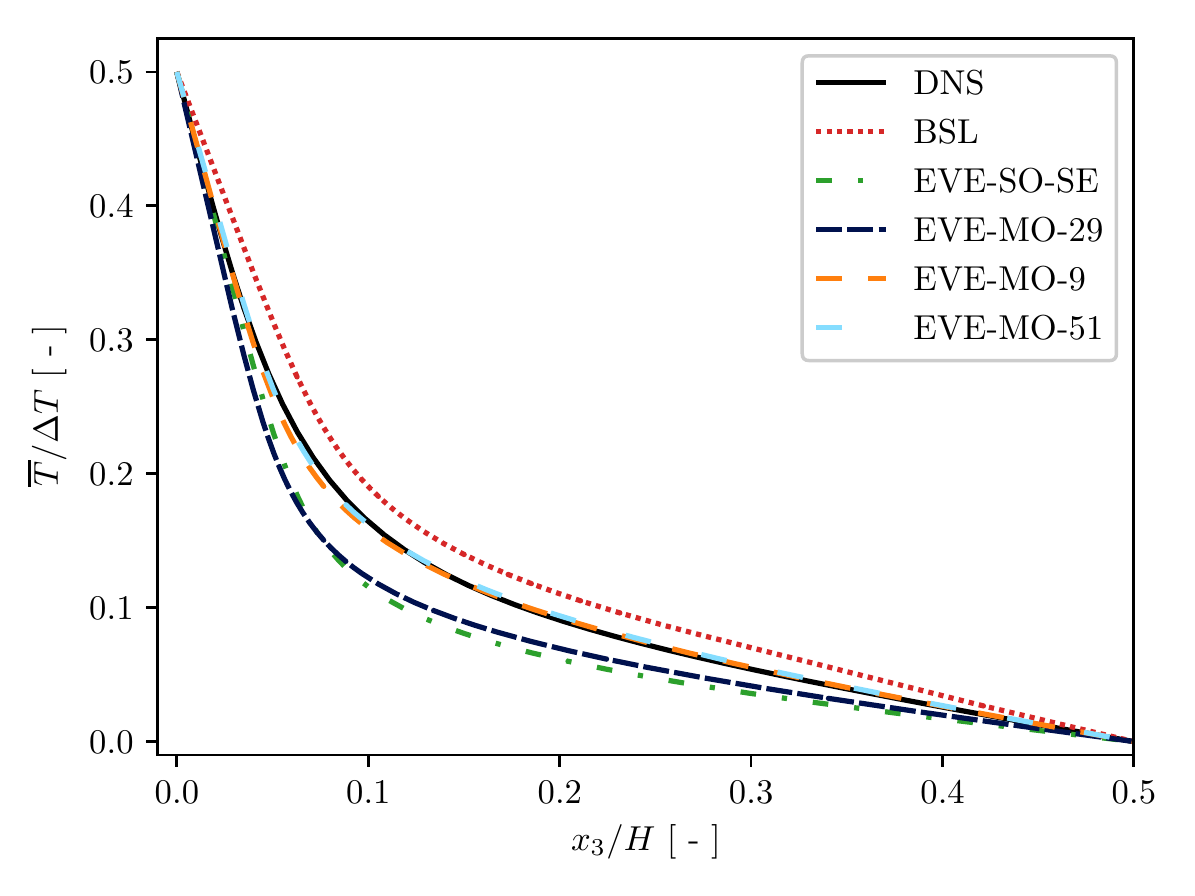}
    \caption{Wall-normal profiles of mean temperature for VNC case at $Ra = 5.4 \times 10^5$.}
    \label{fig:VNC_profiles_T}
\end{figure}

The resulting EVE-MO model predictions of the $\overline{u}_1$ and $\overline{T}$ profiles in the non-dimensionalized wall-normal direction $x_3/H$ are plotted in Figure \ref{fig:VNC_profiles_u} and \ref{fig:VNC_profiles_T}, respectively. The large overprediction of the $\overline{u}_1$ profile by the BSL models is accurately corrected by all three EVE-MO candidate solutions. The EVE-MO-51 models, which produce the highest $J_3$ value, predict a maximum velocity that is slightly too large. In contrast, the individually trained EVE-SO-SE models lead to an underprediction of the mean velocity.

The mean temperature profile in Figure \ref{fig:VNC_profiles_T} is predicted accurately by the EVE-MO-9 and EVE-MO-51 models, while the BSL models significantly underpredict the temperature gradient at the wall and hence, the heat flux into the domain. The EVE-MO-29 models, which do not achieve a sufficient reduction of the training objective $J_4$, as well as the EVE-SO-SE models, predict a temperature gradient that is too large.

\begin{table}[h]
	\centering
	\caption{Scalar functions for $\overline{u^\prime_1 u^\prime_3}$, $R_T$, $\overline{u^\prime_3 T^\prime}$ closure models for VNC case}
    \resizebox{\linewidth}{!}{
		\centering
        \begin{tabular}{l l l l l}
            \toprule
                        & $c(I^1)$ & $d(I^1)$ & $e(I^1, I_T^1)$ \\
            \midrule
            BSL         & $-2.0$ & $0$ & $-1.111$ \\
            EVE-SO-SE   & $- 2.861 + 2.375 \cdot I^1$    & $5.334 - 14.305 \cdot I^1$   & $- 2.909 + 12.281 \cdot I^1 + 1.328 \cdot I_T^1$ \\
                        &                                &                              & $- 5.0 \cdot (I^1)^2 - 5.0 \cdot I^1  \cdot  I_T^1$ \\
            EVE-MO-29   & $- 1.062 + 2.0 \cdot I^1$      & $8.991 + 2.0 \cdot I^1$      & $- 1.313 - 2.0 \cdot I^1 + I_T^1$ \\
            EVE-MO-9    & $- 1.321 + 2.0 \cdot I^1$      & $11.875$                     & $- 0.706 - 2.586 \cdot I^1 + I_T^1$ \\
                        &                                &                              & $+ 1.706 \cdot I^1 \cdot I_T^1$ \\
            EVE-MO-51   & $- 1.336 + 2.0 \cdot I^1$      & $8.916 + 24.663 \cdot I^1$   & $- 0.706 - 2.441 \cdot I^1 + I_T^1$\\
                        &                                & $+ 20.816 \cdot (I^1)^2 + 5.336 \cdot (I^1)^3$ & \\
            \bottomrule
        \end{tabular}
    }
	\label{tab:VNC_expressions}
\end{table}

To further analyze these predictions, we exploit the fact that the EVE framework derives interpretable closure models. Table \ref{tab:VNC_expressions} shows the specific expressions of the selected EVE-MO candidates in contrast to the EVE-SO-SE and BSL models. Considering that the invariants $I^1$ and $I^2$ are of order $\mathcal{O}(10^{-2})$ and $\mathcal{O}(10^{-3})$, respectively, the leading terms of the scalar functions $c$, $d$ and $e$ indicate the approximate directions of modification compared to the BSL models. All four candidate solutions derived by the EVE framework introduce a turbulence production correction via the scalar function $d$. The EVE-SO-SE solution shows the lowest leading term for $d$ with a value of $5.334$ and increases the magnitudes of the functions $c$ and $e$. Since the scalar functions in this training approach are derived independently to correct the shortcomings of the BSL models, we can assume that each of these modifications aims to correct the entire BSL inaccuracies in order to reduce the training objectives $J_3$ and $J_4$. As a consequence, the combination of those independently trained models results in the observed overcorrections for both the $\overline{u}_1$ and $\overline{T}$ profiles. 

\begin{figure}[ht]
    \centering
    \includegraphics[scale=0.75]{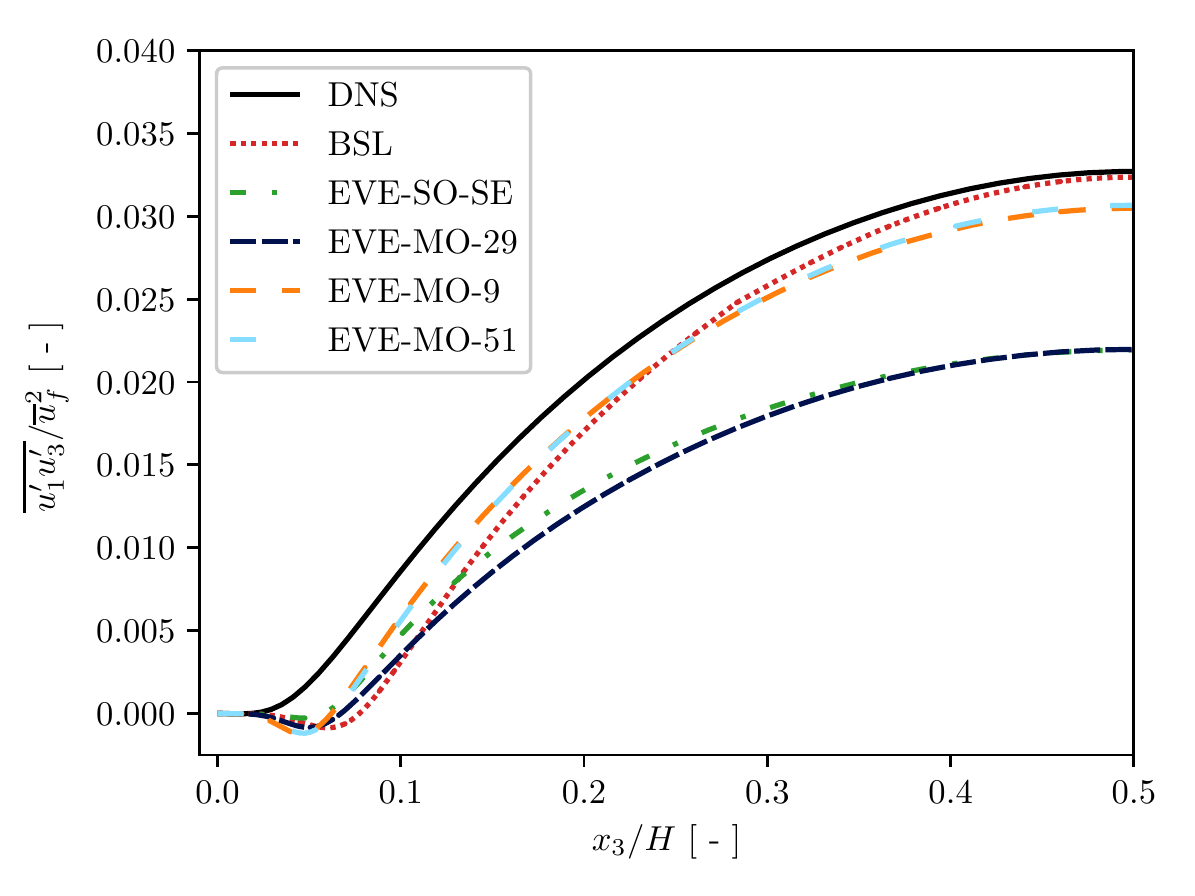}
    \caption{Wall-normal profiles of Reynolds stress for VNC case at $Ra = 5.4 \times 10^5$.}
    \label{fig:VNC_profiles_RS13}
\end{figure}

\begin{figure}[ht]
    \centering
    \includegraphics[scale=0.75]{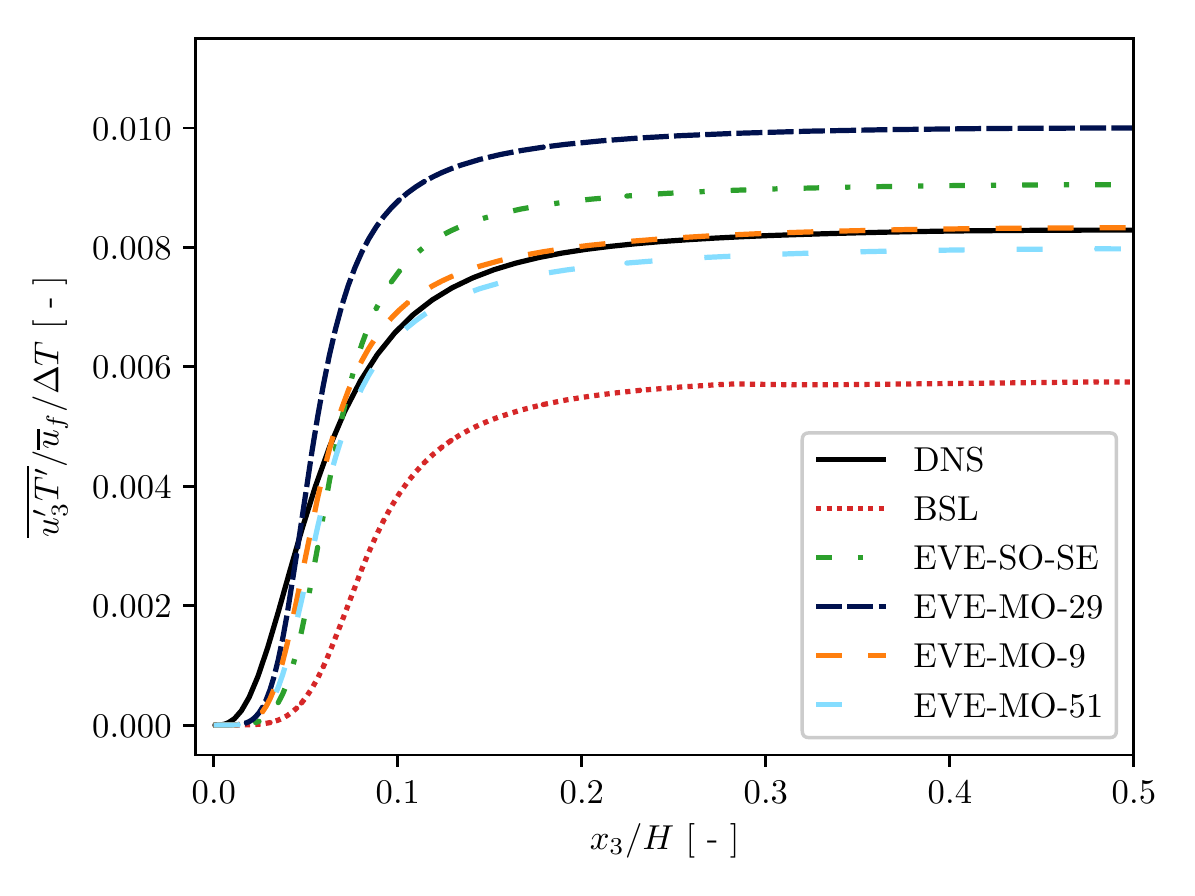}
    \caption{Wall-normal profiles of turbulent heat flux for VNC case at $Ra = 5.4 \times 10^5$.}
    \label{fig:VNC_profiles_HF3}
\end{figure}

Interestingly, the EVE-MO candidates increase the leading terms of the production correction further and mostly decrease the magnitudes of $c$ and $e$, even compared to the BSL models. The EVE-MO-9 and EVE-MO-51 solutions are able to balance the increase in turbulence production and eddy viscosity with maintaining excellent $\overline{u^\prime_1 u^\prime_3}$ and $\overline{u^\prime_3 T^\prime}$ levels (see Figures \ref{fig:VNC_profiles_RS13} and \ref{fig:VNC_profiles_HF3}). In contrast, the EVE-MO-29 candidate fails to decrease the scalar function $e$ magnitude. This leads to an overprediction of the turbulent heat flux and consequently to an inaccurate $\overline{T}$ prediction. The inaccuracy of the mean temperature prediction also influences the $\overline{u}_1$ profile. In order to achieve a good fit to the $\overline{u}_1$ DNS training data, the scalar function $c$ magnitude is reduced too far, which then results in a poor $\overline{u^\prime_1 u^\prime_3}$ prediction.

\begin{figure}[ht]
    \centering
    \includegraphics[scale=0.75]{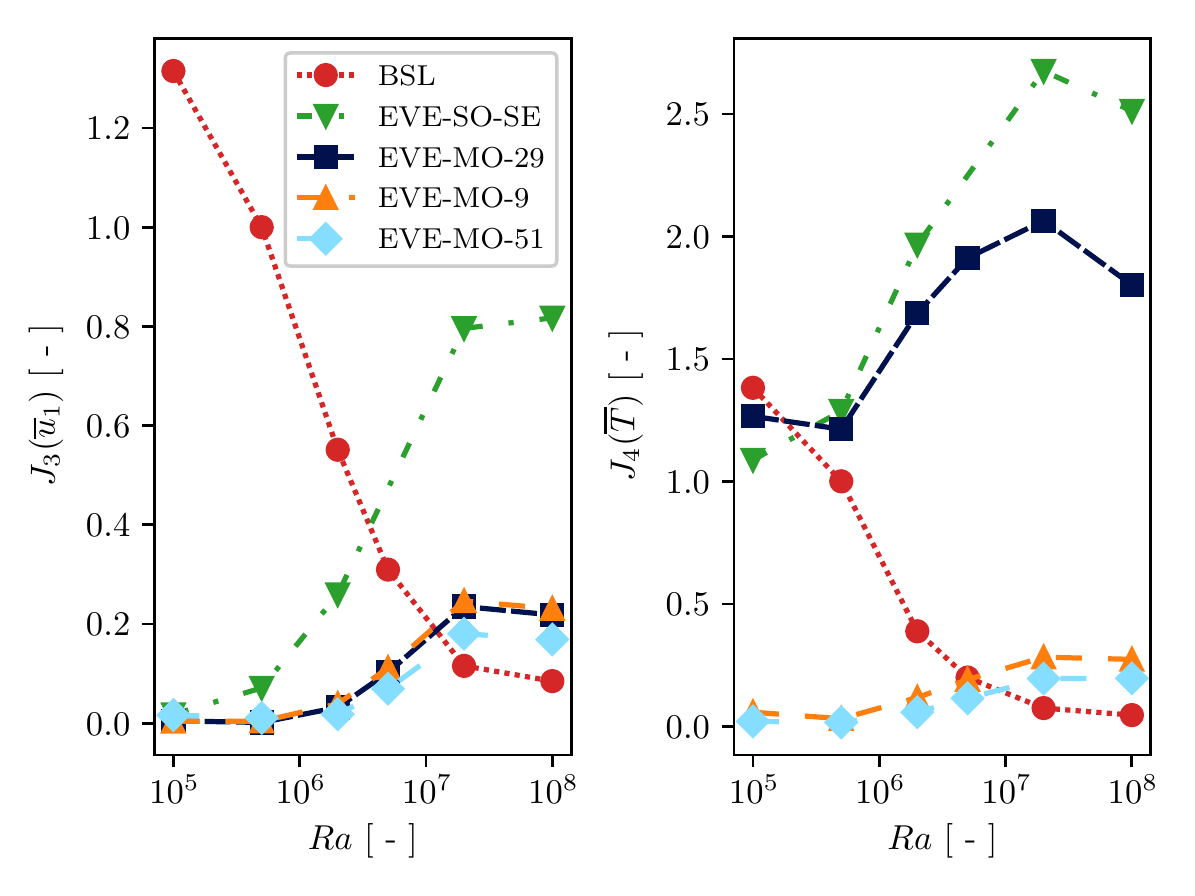}
    \caption{Performance of EVE and BSL models on training objectives $J_3$ (left) and $J_4$ (right) for VNC case at range of Rayleigh numbers.}
    \label{fig:VNC_extrapolation}
\end{figure}

The robustness of the derived EVE candidates is investigated by applying the models trained at $Ra = 5.4 \times 10^5$ to VNC flows at unseen Rayleigh numbers, which were not utilized for training. Figure \ref{fig:VNC_extrapolation} employs the training objectives $J_3$ (left) and $J_4$ (right) to evaluate the mean squared $\overline{u}_1$ and $\overline{T}$ prediction errors of the candidates compared to the DNS data at a range of $Ra$ numbers from $10^5$ to $10^8$. The performance of the BSL models is added for reference.

The EVE-MO-9 and EVE-MO-51 solutions make robust predictions for $\overline{u}_1$ and $\overline{T}$ at the unseen flow conditions and improve over the BSL models up to $Ra = 1 \times 10^7$. At high Rayleigh numbers, the level of turbulence in the near-wall region increases significantly \cite{ng2017}. Figure \ref{fig:VNC_extrapolation} indicates that the BSL models are only calibrated for highly turbulent flow. The two selected solutions, despite being trained at $Ra = 5.4 \times 10^5$, maintain moderate prediction errors at high $Ra$ numbers. As a consequence, the EVE-MO-51 solution shows the most robust $\overline{u}_1$ and $\overline{T}$ predictions across the range of tested $Ra$ numbers. Interestingly, this solution displays the highest robustness on both training objectives, despite yielding the largest $J_3$ value of the selected EVE-MO candidates in training (see Figure \ref{fig:VNC_pareto}). In contrast, the EVE-MO-29 solution leads to large $\overline{T}$ prediction errors at $Ra$ numbers above the training value. Lastly, the individually trained EVE-SO-SE models are not robust, as both the $\overline{u}_1$ and $\overline{T}$ prediction accuracies deteriorate with increasing Rayleigh numbers.

In summary, the EVE-MO training is able to derive models that achieve close-to-ideal predictions of the mean flow quantities $\overline{u}_1$ and $\overline{T}$ for a challenging VNC flow at $Ra = 5.4 \times 10^5$ with highly coupled momentum and thermal fields. The multi-expression training capability is essential to balance the modifications of the trained closure models, as an independent training of the models leads to overcorrections for both the $\overline{u}_1$ and $\overline{T}$ profiles. Due to a diverse pool of candidate solutions created by the multi-objective optimization approach, we are additionally able to select the set of EVE-MO-9 models as a solution that combines excellent reductions of both training objectives with accurate representations of the turbulence quantities and robust predictions across a wide range of $Ra$ numbers. A comparable set of closure models could not be derived with the standard EVE framework.

\section{Conclusion}
\label{sec:conclusion}

Two novel concepts in data-driven turbulence modeling are presented and investigated. These concepts allow the simultaneous training of multiple closure models towards multiple training objectives. Firstly, the standard EVE framework by \citet{weatheritt2016}, which derives algebraic equations from high-fidelity data based on the GEP algorithm, is extended by the multi-expression training capability. The trained closure models have a shared fitness value that is evaluated via a RANS calculation. This allows us to account for the coupled effects of the models on the mean flow quantities. For a vertical natural convection flow with strongly coupled momentum and thermal fields, excellent mean flow quantity predictions were demonstrated that could not be achieved with single-expression training. The training of an anisotropy model $a_{ij}$ and a turbulence production correction $R$ for a periodic hills flow led to improvements of around $10 \si{\percent}$ compared to a state-of-the-art decoupled training approach \cite{schmelzer2019}. This suggests that the coupling effects of the $a_{ij}$ and $R$ models are not strong for this particular flow case.

Secondly, the multi-objective optimization capability is added to the EVE framework, which is based on the NSGA-II algorithm \cite{deb2002}. In contrast to the standard EVE framework, where multiple objectives are combined in a weighted sum approach before running the training, the concept of Pareto dominance is applied. As a consequence, the performance of the derived models is distributed uniformly across the Pareto front instead of focusing on a narrow region. Thus, the user is able to select suitable closure models after inspecting the training results, instead of being required to specify weights beforehand that might or might not lead to good results. The advantages of multi-objective optimization were clearly apparent for both training flows. For the vertical natural convection flow, an analysis of the interpretable model expressions along the Pareto front allowed us to identify closure models where the excellent mean flow predictions were also supported by accurate predictions of the turbulence quantities.

The outcomes of this paper serve as a proof-of-concept for the introduced extensions to the EVE framework. The framework successfully created implementable and interpretable turbulence models for two challenging flow problems. Here, distinct efficacy was observed for a case with a strong coupling of the effects of the trained models on the training objectives. An interesting future application for the novel capabilities is training explicitly coupled closure models, such as a general gradient diffusion hypothesis model in addition to the anisotropy model for a complex heat flux problem.

Furthermore, the robustness of the developed closure models to a wide range of characteristic parameter values was demonstrated for the vertical natural convection case. In order to derive models in the future that perform well even across a range of different flow problems, potentially with varying flow physics, a larger set of diverse training cases will be required. It is the authors' view that the concepts introduced here are important to develop such generalized models, as these turbulence models will likely consist of more than one data-driven closure model and the increasing number of training cases will lead to more competing training objectives that require careful balancing. We further identify the selection of additional input features, which was considered to be a limiting factor to further improvements on the periodic hills case, and a reduction of the computational training costs, which are mainly due to the CFD-driven training approach and will increase with the number of training cases, to be important areas that require future progress to enable the development of generalized data-driven turbulence models.

\section*{Acknowledgments}
This work was supported by a Melbourne Research Scholarship provided by the University of Melbourne.

\appendix
\section{Closure Models for Periodic Hills Flow}
\label{sec:appendix}

\begin{table}[ht]
	\centering
	\caption{Scalar functions in Eq.~\eqref{eq:anisotropyPH} for $a_{ij}$ closure model for PH case}
    \resizebox{\linewidth}{!}{
		\centering
        \begin{tabular}{l l l l}
            \toprule
                        & $g^1(I^1, I^2)$ & $g^2(I^1, I^2)$ & $g^3(I^1, I^2)$ \\
            \midrule
            BSL         & $-2.0$          & $0$             & $0$        \\

            SCH         & $-2.0$          & $0$             & $0$        \\

            EVE-MO-11   & $-1.641 + 4.0 \cdot I^1 + 6.0 \cdot I^2$  & $- 26.18 \cdot I^1 - 4.0 \cdot I^2$      & $8.0 + 6.0 \cdot I^1 + 2.0 \cdot I^2$    \\
                        &                                           & $+ 8.0 \cdot (I^1)^2$                    &                                          \\

            EVE-MO-52   & $-1.641 + 4.0 \cdot I^1 + 4.0 \cdot I^2$  & $1.166 - 30.305 \cdot I^1$               & $16.0 + 4.0 \cdot I^1 - 2.0 \cdot I^2 $  \\
                        &                                           & $+ 8.0 \cdot (I^1)^2$                    &                                          \\

            EVE-MO-57   & $-1.641 + 4.0 \cdot I^1 + 4.0 \cdot I^2$  & $4.0 - 30.0 \cdot I^1$                   & $12.391 + 2.0 \cdot I^1 - 2.0 \cdot I^2$  \\
                        &                                           & $+ 8.0 \cdot (I^1)^2$                    & $+ 2.0 \cdot (I^1)^2$  \\
            
            \bottomrule
        \end{tabular}
    }
	\label{tab:PH_aij_expressions}
\end{table}

\begin{table}[ht]
	\centering
	\caption{Scalar functions in Eq.~\eqref{eq:productionPH} for $R$ closure model for PH case}
    \resizebox{\linewidth}{!}{
		\centering
        \begin{tabular}{l l l l}
            \toprule
                        & $h^1(I^1, I^2)$ & $h^2(I^1, I^2)$ & $h^2(I^1, I^2)$ \\
            \midrule
            BSL         & $0$             & $0$             & $0$  \\

            SCH         & $0.39$          & $0$             & $0$   \\

            EVE-MO-11   & $- 0.806 + 6.0 \cdot I^2$     & $ 4.0 + 1.82 \cdot I^2$          & $5.064 - 2.0 \cdot I^2$          \\
                        &                               & $- 2.0 \cdot (I^1)^2$            &                                  \\

            EVE-MO-52   & $- 1.049 + 6.0 \cdot I^2$     & $- 1.04 - 2.016 \cdot I^1 $      & $6.653 + 2.0 \cdot I^1  + 2.0 \cdot I^2$ \\
                        &                               & $+ 2.0 \cdot (I^1)^2$            & $- 2.0 \cdot (I^2)^2$                     \\

            EVE-MO-57   & $- 1.049 + 6.0 \cdot I^2$     & $- 1.04 \cdot I^1 - 1.04 \cdot I^2 + 0.845 \cdot (I^1)^2 $    & $6.591 + 2.0 \cdot I^1$ \\
                        &                               & $+ 0.016 \cdot (I^1)^3  - 0.016 \cdot I^1 \cdot I^2$        & $- 2.0 \cdot (I^2)^2$  \\
            
            \bottomrule
        \end{tabular}
    }
	\label{tab:PH_R_expressions}
\end{table}

\clearpage
\bibliography{JCP_Manuscript}

\end{document}